\documentclass[prb,twocolumn,superscriptaddress,floatfix,showkeys]{revtex4-1}

\usepackage[version=3]{mhchem} 
\usepackage[dvipsnames]{xcolor}
\usepackage{graphicx}
\usepackage{comment}
\usepackage{amsmath}
\usepackage{soul}
\usepackage{braket}

\definecolor{magenta}{rgb}{1,0,0}

\definecolor{arancione}{rgb}{3,0.6,0.3}

\definecolor{red}{rgb}{1,0,0}

\newcommand{\bq}{\mathbf{q}}
\newcommand{\bn}{\mathbf{n}}
\newcommand{\iz}{i\mathbf{0}}                       
\newcommand{\jq}{j\bq}
\newcommand{\jmq}{j-\bq}
\newcommand{\Qiz}{\hat{Q}_{\iz}}
\newcommand{\Phic}{\Phi_{\iz\jq\jmq}}               

\newcommand{\one}{\mathbf{1}_{\iz}}

\begin{document}
\author{Davide Mitoli}
\affiliation{Dipartimento di Chimica, Universit\`{a} di Torino, via Giuria 5, 10125 Torino, Italy}

\author{Carlo Vadori}
\affiliation{Dipartimento di Chimica, Universit\`{a} di Torino, via Giuria 5, 10125 Torino, Italy}

\author{Kim Eklund}
\affiliation{Dipartimento di Chimica, Universit\`{a} di Torino, via Giuria 5, 10125 Torino, Italy}
\affiliation{Department of Chemistry and Materials Science, Aalto University, Kemistintie 1, FI-00076 Aalto, Finland}

\author{Jefferson Maul}
\affiliation{Dipartimento di Chimica, Universit\`{a} di Torino, via Giuria 5, 10125 Torino, Italy}
\affiliation{Academic Directorate of Sciences (DIACC), Natal-Central Campus (CNAT), Federal Institute of Education, Science and Technology of Rio Grande do Norte (IFRN), Av. Senador Salgado Filho, 1559, Tirol, Natal, RN, 59015-000, Brazil}

\author{Antti J. Karttunen}
\affiliation{Department of Chemistry and Materials Science, Aalto University, Kemistintie 1, FI-00076 Aalto, Finland}

\author{Alessandro Erba}
\email{alessandro.erba@unito.it}
\affiliation{Dipartimento di Chimica, Universit\`{a} di Torino, via Giuria 5, 10125 Torino, Italy}

\date{\today}

\title{Anharmonic Phonon Theory of Primary Pyroelectricity}

\begin{abstract}

We present a periodic extension of vibrational perturbation theory (p-VPT) for the first-principles evaluation of the primary (clamped-lattice) pyroelectric effect in polar crystals. This perturbative approach allows for a systematic assessment of the effects of anharmonic phonon-phonon couplings at a modest cost compared to traditional variational methods such as vibrational self-consistent field (VSCF) and vibrational configuration interaction (VCI). This efficiency stems from two key factors. Firstly, only odd-order potential terms enter our p-VPT expressions; in contrast, variational truncations at odd orders may lead to unbounded total or mean-field potentials, triggering severe numerical instabilities. Secondly, the compact working expressions for finite-temperature atomic displacements and primary pyroelectric coefficients bypass the need to explicitly sample the vibrational Fock space. To address this sampling challenge in VSCF, we have developed a stochastic Metropolis-Hastings approach. Using bulk wurtzite-structured ZnO as a rigorous benchmark system, we show that p-VPT achieves remarkable quantitative agreement with variational calculations, across local, gradient-corrected, and hybrid functionals. Notably, we find that  a single phonon approximation (SPA), which isolates the anharmonicity of the single pyroelectrically active \(A_1(\text{TO})\) mode at \(\Gamma \) (i.e. neglecting phonon-phonon couplings and the intrinsic anharmonicity of all other modes), provides a qualitatively sound description of the primary pyroelectric coefficient across a wide temperature range. 
At a minimal computational cost, the SPA may offer a viable route for high-throughput pyroelectric materials screening, while full quantitative accuracy is reproduced within p-VPT by explicitly accounting for phonon dressing.




\end{abstract}
\maketitle
\section{Introduction}

In the pyroelectric effect, temperature fluctuations induce a change in the spontaneous polarization of a material.\cite{whatmore1986pyroelectric,lang2005pyroelectricity} A necessary condition for spontaneous polarization, and thus pyroelectric behavior, is that the material crystallizes in a polar space group.\cite{nye1985physical} The pyroelectric coefficient is defined as the rate of change of polarization with respect to temperature, and it can be decomposed into distinct physical contributions. The primary contribution, often termed the ``true'' pyroelectric effect, occurs at constant strain and is also referred to as ``fixed-volume'' or ``clamped-lattice'' pyroelectricity. The secondary pyroelectric effect arises from thermal expansion and the resulting piezoelectric strain; consequently, its coefficient can be directly calculated from thermal expansion coefficients and piezoelectric constants.\cite{bhalla1980primary} A third contribution, the tertiary pyroelectric effect, originates from non-uniform heating and the resulting thermal stress, making it dependent on sample geometry and size.\cite{whatmore1986pyroelectric,kosorotov1994dynamic} Typically, but not always, the primary coefficient is the largest of the contributions. Pyroelectric materials are well established in sensor and detector technologies,\cite{whatmore1986pyroelectric,whatmore2023pyroelectric} with more recent applications emerging in catalysis\cite{schlechtweg2019novel,li2020recent} and energy harvesting.\cite{bowen2014pyroelectric,zhang2020thermal,hanrahan2022pyroelectrics} 

Pioneering first-principles modeling of pyroelectricity was reported by Peng and Cohen for ferroelectric \ce{LiNbO3} using a shell-model potential in molecular dynamics (MD) simulations.\cite{peng2011origin} \textit{Ab initio} MD (AIMD) has since been applied to study pyroelectricity in other ferroelectric systems, including perovskite \ce{PbTiO3},\cite{lynch2022insight} \ce{HfO2},\cite{liu2019origin,liu2026design} \ce{ZrO2},\cite{ganser2022piezo} and monolayer \ce{In2Se3}.\cite{liu2019pyroelectric} Approaches based on the density-functional theory (DFT) to pyroelectricity in ferroelectric oxides include calculations using experimental atomic displacements for \ce{LiNbO3} and \ce{LiTaO3} by Weigel \textit{et al.},\cite{weigel2023spontaneous} as well as the work of Eklund and Karttunen on perovskite ferroelectrics \ce{BaTiO3},\cite{eklund2023pyroelectric} \ce{KNbO3},\cite{eklund2023pyroelectric} \ce{PbTiO3},\cite{eklund2024pyroelectricity} and ordered models of the solid solutions \ce{Pb(Zr_{0.5}Ti_{0.5})O3}\cite{eklund2025pyroelectric} and \ce{(K_{0.5}Na_{0.5})NbO3}.\cite{eklund2025pyroelectric_knn} The latter methodology employs self-consistent phonon (SCPH) calculations to obtain anharmonic finite-temperature phonon properties, which are then used to generate thermally displaced atomic configurations. The primary pyroelectric coefficient is subsequently evaluated via the Berry-phase approach using a central finite difference between two finite-temperature structural models.

Owing to their simple non-centrosymmetric, polar, structure, wurtzite-type materials serve as an ideal benchmark for assessing the predictive capability of density functional theory (DFT) in capturing the subtle electronic and structural nuances that govern spontaneous polarization and related electromechanical and thermal responses, such as piezoelectricity and pyroelectricity.\cite{dal1994ab,massidda1995polarization,hill2000first,catti2003full,wu2005systematic, liu2018internal,rostami2025anisotropic} In this context, several quasi-harmonic methodologies have been developed to describe the pyroelectric behavior of prototype wurtzite systems like ZnO and GaN. For instance, Liu {\it et al.} investigated both primary and secondary pyroelectricity in these compounds via the zero static internal stress approximation (ZSISA).\cite{liu2016first,liu2018mechanisms} Later, Masuki {\it et al.} performed full finite-temperature optimizations of the unit cell and atomic coordinates, minimizing the Helmholtz free energy to determine the internal degrees of freedom.\cite{masuki2023full} More recently, Gong and Dal Corso evaluated the performance of both ZSISA and full free energy minimization (FFEM) within the quasi-harmonic approximation (QHA) for ZnO.\cite{gong2026ab} 




This work introduces an alternative first-principles framework to primary pyroelectricity in polar materials based on a periodic extension of vibrational perturbation theory (p-VPT), building on previous work done in both a molecular\cite{biczysko2010harmonic,barone2012toward,bloino2012second,barone2014fully} and condensed-matter\cite{PhysRev.128.2589,calandra2007anharmonic,PhysRevB.82.104504,PhysRevB.86.155125,paulatto2015first} context. By exploiting compact analytical expressions, this scheme captures the intricate effects of anharmonic phonon-phonon interactions without the heavy computational burden~\cite{PRL_RESO} or potential numerical instabilities~\cite{yang2023pitfalls,spencer2026converging} often encountered in traditional variational treatments like vibrational self-consistent field (VSCF) and vibrational configuration interaction (VCI). To establish a rigorous baseline for verification, we also develop a stochastic Metropolis-Hastings sampling technique to tackle the combinatorial complexity of the vibrational Fock space within the VSCF variational approach.

We have implemented the new methodology in a developmental version of the \textsc{Crystal} program~\cite{Erba2023} for electronic structure theory of solids. Using bulk wurtzite ZnO as a benchmark system, we demonstrate that p-VPT achieves excellent quantitative agreement with variational methods across local, gradient-corrected, and hybrid density functionals. Crucially, we introduce and evaluate a single phonon approximation (SPA) that isolates the anharmonicity of the single pyroelectrically active mode at $\Gamma$, suggesting its utility as a computationally inexpensive tool for high-throughput materials screening. Finally, we show how full quantitative accuracy is systematically recovered within p-VPT by accounting for phonon dressing through dense Brillouin zone sampling.

\section{Theory}
\label{sec:theory}

We describe lattice dynamics within the Born-Oppenheimer approximation. States associated with vibrational degrees of freedom are thus obtained by solving the nuclear Schr\"odinger equation:
\begin{equation}
\label{eq:nucSchro}
\hat{H} \vert s \rangle = E_s \vert s \rangle \quad \textup{with} \quad \hat{H} = \hat{T} + \hat{V} \; ,
\end{equation}
where $\hat{T}$ is the kinetic energy operator for the nuclei in the lattice and $\hat{V}$ is the potential energy surface (PES) describing the energy of the electronic system as a function of the nuclear coordinates. We compute the PES using  DFT with exchange-correlation (xc) functionals from different levels of approximation: local density (LDA), generalized-gradient (GGA), and hybrid GGA.

\subsection{Harmonic Phonons}
\label{sec:harm}

We evaluate the harmonic phonons of a periodic lattice via the direct frozen-phonon (FP) approach (also referred to as the finite-displacement or small-displacement method),\cite{wei1992ab,PhysRevLett.78.4063,PhysRevB.61.272,van2002effect,alfe2009phon,togo2015first} in which second-order interatomic force constants (IFCs) are computed by displacing atoms in the reference primitive cell and analytically evaluating the resulting forces acting on all atoms within a supercell (SC):\cite{PyrGro}
\begin{equation}
\label{eq:qifcs}
  \Phi^\mathbf{g}_{a\alpha b\beta} = \left.\frac{\partial^2 E}{\partial u_{a\alpha}^{\mathbf{0}}\partial u_{b\beta}^\mathbf{g}}\right\vert_\textup{eq} \equiv  \Phi^\mathbf{g}_{ij} = \left.\frac{\partial^2 E}{\partial u_i^{\mathbf{0}}\partial u_j^\mathbf{g}}\right\vert_\textup{eq} \; ,
\end{equation}
where $\mathbf{0}$ labels the reference primitive cell, $\mathbf{g}$ identifies any direct-space lattice vector within the SC, $a,b = 1, \dots, N$ are atomic indices, $\alpha, \beta = x,y,z$ are Cartesian indices, and $i,j = 1, \dots, 3N$ are combined atomic and Cartesian indices labeling all possible Cartesian displacements $\mathbf{u}$ of the $N$ atoms in the primitive cell. 

Dynamical matrices $\mathbf{W}^{\mathbf{q}}$ at wave vectors $\mathbf{q}$ within the first Brillouin zone are obtained via a Fourier transform of the mass-weighted IFCs from Eq.~(\ref{eq:qifcs}):
\begin{equation}
    \mathbf{W}^{\mathbf{q}} = \sum_{\mathbf{g} \in \mathrm{SC}}\mathbf{M}^{-\frac{1}{2}}\boldsymbol{\Phi}^{\mathbf{g}}\mathbf{M}^{-\frac{1}{2}} e^{i\mathbf{q\cdot g}} \; ,
    \label{eq:dyn_mat}
\end{equation}
where $\mathbf{M}$ is the $3N\times 3N$ diagonal atomic mass matrix. Diagonalization of the dynamical matrices in Eq.~(\ref{eq:dyn_mat}),
\begin{equation}
    \label{eq:diag}
    \mathbf{W}^{\mathbf{q}}  \mathbf{U}^{\mathbf{q}}  = \boldsymbol{\Lambda}^{\mathbf{q}}  \mathbf{U}^{\mathbf{q}} \; ,  
\end{equation}
yields the harmonic phonon frequencies $\omega_{i\mathbf{q}} = \lambda_{i\mathbf{q}}^{\frac{1}{2}}$ and the associated normal-mode coordinates $Q_{i\mathbf{q}}$ in the basis of mass-weighted Cartesian displacements:
\begin{equation}
\label{eq:Q}
\mathbf{Q}^{\mathbf{q}}=(\mathbf{U}^{\mathbf{q}})^\dagger \mathbf{M}^{\frac{1}{2}}\mathbf{u}^{\mathbf{q}} \quad \textup{with} \quad \mathbf{u}^{\mathbf{q}}= \frac{1}{\sqrt{N_g}}\sum_{\mathbf{g} \in \mathrm{SC}} \mathbf{u}^{\mathbf{g}} e^{i\mathbf{q} \cdot \mathbf{g}} \; ,
\end{equation}
where $N_g$ is the number of primitive cells within the SC. At the harmonic level, all phonons are independent. In the basis of the phonon normal coordinates, the nuclear Hamiltonian operator of Eq.~(\ref{eq:nucSchro}) takes the harmonic form:
\begin{equation}
\label{eq:H0}
\hat{H}^{(0)} = \sum_{i\mathbf{q}} \hat{H}^{(0)}_{i\mathbf{q}} \quad \textup{with} \quad \hat{H}^{(0)}_{i\mathbf{q}} = \frac{1}{2} \left[-\frac{\partial^2}{\partial Q_{i\mathbf{q}}^2} + \omega_{i\mathbf{q}}^2 \hat{Q}_{i\mathbf{q}}^2 \right] \; ,
\end{equation}
where each phonon mode has associated energy levels and eigenstates satisfying $\hat{H}^{(0)}_{i\mathbf{q}} \vert n_{i\mathbf{q}}^{(0)}\rangle = \varepsilon_{i\mathbf{q},n_{i\mathbf{q}}}^{(0)} \vert n_{i\mathbf{q}}^{(0)}\rangle$, with $n_{i\mathbf{q}} = 0, 1, 2, \dots$ denoting the number of excitation quanta in each mode and $\varepsilon_{i\mathbf{q},n_{i\mathbf{q}}}^{(0)} = (\frac{1}{2} + n_{i\mathbf{q}})\hbar \omega_{i\mathbf{q}}$. The total harmonic nuclear Hamiltonian $\hat{H}^{(0)}$ of Eq.~(\ref{eq:H0}) satisfies the eigenvalue equation: 
\begin{equation}
\label{eq:h0psi0}
\hat{H}^{(0)} \vert \mathbf{n}^{(0)} \rangle = E_{\mathbf{n}}^{(0)} \vert \mathbf{n}^{(0)} \rangle \; .
\end{equation} 
At the harmonic level, the vibrational states can be labeled by configuration vectors $\mathbf{n} = (\cdots, n_{i\mathbf{q}}, \cdots)$ collecting the occupation numbers of all phonon modes. These states take the product form $\vert s^{(0)} \rangle \equiv \vert \mathbf{n}^{(0)} \rangle = \bigotimes_{i\mathbf{q}} \vert {n_{i\mathbf{q}}}^{(0)}\rangle$, with total energies $E_s^{(0)} \equiv E_{\mathbf{n}}^{(0)} = \sum_{i\mathbf{q}} \varepsilon_{i\mathbf{q},n_{i\mathbf{q}}}^{(0)}$.

The direct approach outlined above ensures that the computed harmonic phonons are exact at wave vectors $\mathbf{q}$ that are \textit{commensurate} with the SC. Phonons at arbitrary wave vectors $\mathbf{q}$ can be interpolated via Eqs.~(\ref{eq:dyn_mat})--(\ref{eq:diag}) only if the IFCs in Eq.~(\ref{eq:qifcs}) decay to zero within the spatial extent of the SC. For nonpolar materials, this condition can be met by choosing sufficiently large SCs. Polar materials, however, require an explicit treatment of long-range electrostatic interactions, which give rise to the splitting between longitudinal and transverse optical phonon frequencies (LO--TO splitting).\cite{gonze2005first}

\subsubsection{Non-analytical Long-Range Correction}
\label{sec:long}

To account for long-range electrostatic interactions in the phonon dispersion of polar materials, we adopt the mixed-space approach introduced by Wang \textit{et al.},\cite{wang2010mixed} in which the IFC matrix $\boldsymbol{\Phi}^{\mathbf{g}}$ in Eq.~(\ref{eq:dyn_mat}) is replaced by $\tilde{\boldsymbol{\Phi}}^{\mathbf{g}}$:
\begin{equation}
\label{eq:Phitilde}
 \tilde{\Phi}^\mathbf{g}_{a\alpha b\beta}  = \Phi^\mathbf{g}_{a\alpha b\beta} + \Xi_{a\alpha b\beta}^{\mathrm{na}} \; ,
\end{equation}
where a nonanalytical (na) long-range term $\Xi_{a\alpha b\beta}^{\mathrm{na}}$ is added to the short-range IFCs $\Phi^\mathbf{g}_{a\alpha b\beta}$ of Eq.~(\ref{eq:qifcs}):
\begin{equation}
\label{wang}
\Xi_{a\alpha b\beta}^{\mathrm{na}} = \frac{1}{N_g} \frac{4\pi e^2}{V} \frac{[\mathbf{q} \cdot \mathbf{Z}^{\ast}_{a}]_\alpha [\mathbf{q} \cdot \mathbf{Z}^{\ast}_{b}]_\beta}{\mathbf{q} \cdot \boldsymbol{\epsilon}_{\infty} \cdot \mathbf{q}} \; ,
\end{equation}
where $V$ is the primitive cell volume, $e$ is the elementary charge, $\boldsymbol{\epsilon}_{\infty}$ is the high-frequency (clamped-ion) dielectric tensor, and $\mathbf{Z}^{\ast}_{a}$ is the Born effective charge tensor of atom $a$. We compute these dielectric quantities using density functional perturbation theory (DFPT) as described in Refs.~\onlinecite{ferrero2008coupled,maschio2012ab}.

\subsection{Anharmonic Phonons}
\label{sec:anharm}

We expand the potential energy operator in Eq.~(\ref{eq:nucSchro}) in a Taylor series in the basis of the harmonic normal-mode coordinates as follows:
\begin{align}
\label{PesTaylor0}
\hat{V} &= \frac{1}{2}\sum_{i\mathbf{q}} \omega_{i\mathbf{q}}^2 \hat{Q}_{i\mathbf{q}}^2  \\
\label{PesTaylor1}
&+ \frac{1}{3!}\sum_{i\mathbf{q}}\sum_{j\mathbf{q}'}\sum_{k\mathbf{q}''}  
\Phi_{i\mathbf{q}j\mathbf{q}' k\mathbf{q}''}
\hat{Q}_{i\mathbf{q}} \hat{Q}_{j\mathbf{q}'} \hat{Q}_{k\mathbf{q}''}  \\
\label{PesTaylor2}
&+ \frac{1}{4!}\sum_{i\mathbf{q}}\sum_{j\mathbf{q}'}\sum_{k\mathbf{q}''}  \sum_{l\mathbf{q}'''}
\Phi_{i\mathbf{q}j\mathbf{q}' k\mathbf{q}'' l\mathbf{q}'''}
\hat{Q}_{i\mathbf{q}} \hat{Q}_{j\mathbf{q}'} \hat{Q}_{k\mathbf{q}''} \hat{Q}_{l\mathbf{q}'''} \\
&+ \cdots \; , \nonumber
\end{align}
where the third- and fourth-order anharmonic force constants are defined as:
\begin{align}
\label{Phi3}
  \Phi_{i\mathbf{q}j\mathbf{q}' k\mathbf{q}''} &= \left.\frac{\partial^3 E}{\partial Q_{i\mathbf{q}} \partial Q_{j\mathbf{q}'}\partial Q_{k\mathbf{q}''}}\right\vert_\textup{eq} \; ,\\
\label{Phi4}
  \Phi_{i\mathbf{q}j\mathbf{q}' k\mathbf{q}'' l\mathbf{q}'''} &= \left.\frac{\partial^4 E}{\partial Q_{i\mathbf{q}} \partial Q_{j\mathbf{q}'}\partial Q_{k\mathbf{q}''} \partial Q_{l\mathbf{q}'''}}\right\vert_\textup{eq} \; .
\end{align}
Anharmonic phonon states are obtained by solving Eq.~(\ref{eq:nucSchro}) in the presence of higher-than-quadratic terms in the potential. Below, we outline the different approaches employed in this work to solve this problem.  

\subsubsection{Vibrational Perturbation Theory}
\label{sec:vpt2}

Building on the second-order vibrational perturbation theory (VPT2) developed in a molecular quantum-chemical context by Barone and collaborators,\cite{biczysko2010harmonic,barone2012toward,bloino2012second,barone2014fully} we introduce a periodic extension of the method, p-VPT, by partitioning the nuclear Hamiltonian operator of Eq.~(\ref{eq:nucSchro}) into three terms:
\begin{equation}
\label{eq:Hpertu}
\hat{H} = \hat{H}^{(0)} + \lambda \hat{H}^{(1)} + \lambda^2 \hat{H}^{(2)} \; , 
\end{equation}
where $\hat{H}^{(0)}$ is the unperturbed harmonic Hamiltonian from Eq.~(\ref{eq:H0}) with known solutions from Eq.~(\ref{eq:h0psi0}), $\lambda$ is the standard dimensionless perturbation parameter, and the first- and second-order perturbative operators $\hat{H}^{(1)}$ and $\hat{H}^{(2)}$ correspond to the cubic and quartic terms in Eqs.~(\ref{PesTaylor1}) and (\ref{PesTaylor2}), respectively. Both the eigenstates of $\hat{H}$ and their associated energies are expanded in a perturbation series to second order as:
\begin{align}
\label{eq:perts}
  \vert \mathbf{n} \rangle &= \vert \mathbf{n}^{(0)} \rangle + \lambda \vert \mathbf{n}^{(1)} \rangle + \lambda^2 \vert \mathbf{n}^{(2)} \rangle \; ,\\
  \label{eq:perte}
  E_{\mathbf{n}} &= E_{\mathbf{n}}^{(0)} + \lambda E_{\mathbf{n}}^{(1)} + \lambda^2 E_{\mathbf{n}}^{(2)} \; .
\end{align}
Within nondegenerate Rayleigh--Schr\"odinger perturbation theory, the first- and second-order energy corrections are given by:
\begin{align}
\label{eq:e1}
    E_\mathbf{n}^{(1)}  &= \braket{\mathbf{n}^{(0)} | \hat{H}^{(1)} | \mathbf{n}^{(0)}} \; , \\ 
    E^{(2)}_\mathbf{n} 
    \label{eq:e2}
    &= \braket{\mathbf{n}^{(0)} | \hat{H}^{(2)}| \mathbf{n}^{(0)}} + 
    \sum_\mathbf{m}{}^{'} \frac{|\braket{\mathbf{n}^{(0)} | \hat{H}^{(1)} |
    \mathbf{m}^{(0)}} |^2} {E_\mathbf{n}^{(0)} - E_\mathbf{m}^{(0)}} \; ,
\end{align}
where the primed summation in Eq.~(\ref{eq:e2}) runs over all vibrational configurations $\mathbf{m}$ satisfying $E_\mathbf{m}^{(0)} \neq E_\mathbf{n}^{(0)}$. Correspondingly, the first- and second-order state corrections are:
\begin{align}
\label{eq:n1}
    \ket{\mathbf{n}^{(1)}} &= \sum_\mathbf{m}{}^{'} \frac{\braket{\mathbf{n}^{(0)} | \hat{H}^{(1)} | \mathbf{m}^{(0)}}} {E_\mathbf{n}^{(0)} - E_\mathbf{m}^{(0)}} \ket{\mathbf{m}^{(0)}} \; , \\
    \ket{\mathbf{n}^{(2)}} &= \sum_\mathbf{m}{}^{'} \left[\frac{\braket{\mathbf{n}^{(0)} | \hat{H}^{(2)} | \mathbf{m}^{(0)}}} 
    {E_\mathbf{n}^{(0)} - E_\mathbf{m}^{(0)}} \right.\nonumber \\
    \label{eq:n2}
    &+ \left. \sum_\mathbf{l}{}^{'} 
        \frac{\braket{\mathbf{m}^{(0)} | \hat{H}^{(1)}| \mathbf{l}^{(0)}}
        \braket{\mathbf{n}^{(0)} | \hat{H}^{(1)}| \mathbf{l}^{(0)}}} 
    {(E_\mathbf{n}^{(0)} - E_\mathbf{m}^{(0)})
    (E_\mathbf{n}^{(0)} - E_\mathbf{l}^{(0)})}\right]
    \ket{\mathbf{m}^{(0)}} \; .
\end{align}
Explicit analytical expressions for the matrix elements appearing in Eqs.~(\ref{eq:e1})--(\ref{eq:n2}) can be derived by exploiting the properties of the harmonic oscillator basis functions. We detail this derivation in Sec.~\ref{sec:pyropt} and Appendix~\ref{app:pt12} for the specific case of the primary pyroelectric coefficient. Here, we simply note that the first-order energy correction vanishes identically, i.e., $E_\mathbf{n}^{(1)} = 0$.

\subsubsection{Variational Approaches}
\label{sec:vscftheo}

\paragraph{Vibrational Self-Consistent Field.}
Within the vibrational self-consistent field (VSCF) approach, originally formulated in a molecular context\cite{bowman1978self,bowman1986self} and later extended to periodic systems,\cite{monserrat2013anharmonic,PARTII_ANHARM} each vibrational configuration state is expressed as a Hartree product of orthonormal single-phonon states:
\begin{equation}
\label{eq:states}
\vert \mathbf{n} \rangle = \bigotimes_{i\mathbf{q}} \vert n_{i\mathbf{q}} \rangle \; .
\end{equation}
VSCF consists of variationally determining the optimal single-phonon states that minimize the energy $E_{\mathbf{n}} = \langle \mathbf{n} \vert \hat{H}\vert \mathbf{n}\rangle$. For each vibrational configuration $\vert \mathbf{n} \rangle$, this variational condition leads to a set of effective single-phonon equations:
\begin{equation}
\label{eq:speq}
    \hat{H}_{\mathbf{n}, i\mathbf{q}} \vert n_{i\mathbf{q}} \rangle = \varepsilon_{i\mathbf{q},n_{i\mathbf{q}}}\vert n_{i\mathbf{q}} \rangle \; ,
\end{equation}
with $\hat{H}_{\mathbf{n}, i\mathbf{q}} = -\frac{1}{2}\frac{\partial^2}{\partial Q_{i\mathbf{q}}^2} + \hat{V}_{i\mathbf{q}}$, where the mean-field effective single-phonon potential is defined as:
\begin{equation}
    \hat{V}_{i\mathbf{q}} = \left\langle {\bf n}^\prime \middle\vert \hat{V} \middle\vert  {\bf n}^\prime\right\rangle \; ,
\end{equation}
with $\vert {\bf n}^\prime \rangle = \bigotimes_{j{\bf q}^\prime}^\prime\vert {n_{j{\bf q}^\prime}} \rangle$, where the primed product excludes the mode $j\mathbf{q}' = i\mathbf{q}$. The total energy of each state is then obtained from:
\begin{align}
E_{\mathbf{n}} &= \langle \mathbf{n} \vert \hat{H}\vert \mathbf{n}\rangle \nonumber \\
\label{eq:vscfene}
&= \sum_{i\mathbf{q}} \varepsilon_{i\mathbf{q},n_{i\mathbf{q}}} + \left\langle \mathbf{n} \middle\vert \hat{V} - \sum_{i\mathbf{q}} \hat{V}_{i\mathbf{q}} \middle\vert \mathbf{n} \right\rangle \; .
\end{align}
By expanding each single-phonon state in a basis of $M$ harmonic oscillator states,
\begin{equation}
\label{eq:linVSCF}
   \vert n_{i\mathbf{q}} \rangle = \sum_{m_{i\mathbf{q}} = 0}^{M-1} C_{m_{i\mathbf{q}},n_{i\mathbf{q}}}\vert m_{i\mathbf{q}}^{(0)}\rangle \; ,
\end{equation}
the single-phonon equations in Eq.~(\ref{eq:speq}) are cast into algebraic eigenvalue problems that can be solved via matrix diagonalization for each vibrational configuration $\mathbf{n}$ and each phonon mode $i\mathbf{q}$:
\begin{equation}
\mathbf{H}^{\mathbf{n}, i\mathbf{q}} \mathbf{C}^{\mathbf{n}, i\mathbf{q}} = \mathbf{C}^{\mathbf{n}, i\mathbf{q}} \boldsymbol{\varepsilon}^{\mathbf{n}, i\mathbf{q}} \; .
\end{equation}
Here, the effective single-phonon Hamiltonian matrix has elements $H^{\mathbf{n}, i\mathbf{q}}_{lm} = \langle l^{(0)}_{i\mathbf{q}}\vert \hat{H}_{\mathbf{n}, i\mathbf{q}}\vert m^{(0)}_{i\mathbf{q}}\rangle$, the eigenvector matrix $\mathbf{C}^{\mathbf{n}, i\mathbf{q}}$ contains the expansion coefficients of Eq.~(\ref{eq:linVSCF}), and $\boldsymbol{\varepsilon}^{\mathbf{n}, i\mathbf{q}}$ is the diagonal matrix of eigenvalues. 

\paragraph{Vibrational Configuration Interaction.} 
In the vibrational configuration interaction (VCI) approach, originally proposed in a molecular context\cite{bowman1979application} and later extended to periodic systems,\cite{PARTII_ANHARM,PRL_RESO} Eq.~(\ref{eq:nucSchro}) is solved by expressing each anharmonic vibrational state $\vert s \rangle$ as a linear combination of the configuration states from Eq.~(\ref{eq:states}):
\begin{equation}
\label{eq:vcifunc}
\vert s \rangle = \sum_{\mathbf{n}} A_{\mathbf{n},s} \vert \mathbf{n} \rangle \; ,
\end{equation} 
where the single-phonon states $\vert n_{i\mathbf{q}} \rangle$ can either be chosen as the harmonic oscillator functions, $\vert n_{i\mathbf{q}} \rangle = \vert n_{i\mathbf{q}}^{(0)}\rangle$ (VCI-HO), or obtained from a VSCF calculation for a reference state (typically the ground state) according to Eq.~(\ref{eq:linVSCF}) (VCI-VSCF). The method involves constructing the Hamiltonian matrix $\mathbf{H}$ in the basis of configuration states -- with matrix elements $H_{\mathbf{n}\mathbf{m}} = \langle \mathbf{n} \vert \hat{H}\vert \mathbf{m}\rangle$ -- followed by its diagonalization, $\mathbf{H} \mathbf{A} = \mathbf{A} \mathbf{E}$, where $\mathbf{A}$ is the matrix containing the expansion coefficients $A_{\mathbf{n},s}$ of Eq.~(\ref{eq:vcifunc}) column-wise, and $\mathbf{E}$ is the diagonal matrix of the corresponding eigenvalues $E_s$. The expansion in Eq.~(\ref{eq:vcifunc}) must be carefully truncated; practical truncation schemes are discussed in detail elsewhere.\cite{PARTII_ANHARM,schireman2022anharmonic,mitoli2024first,PRL_RESO}

\subsection{Pyroelectricity}

The pyroelectric effect describes the ability of certain crystals to change their spontaneous polarization, ${\bf P}_\textup{S}$, with respect to variations in temperature, $T$, at constant stress, $\sigma$.\cite{lang1974pyroelectricity,lang2005pyroelectricity} According to the Born-Szigeti quantum theory of pyroelectricity,\cite{born1945quantum,szigeti1975temperature} the total pyroelectric coefficient, ${\bf p}(T)$, is commonly expressed as the sum of a primary, ${\bf p}^\prime(T)$, and a secondary, ${\bf p}^{\prime\prime}(T)$, term as follows:
\begin{eqnarray}
   \label{eq:pyrocoeff}
    {\bf p}(T) = \left( \frac{d{\bf P}_\textup{S}}{dT}\right)_\sigma &=& \left( \frac{\partial{\bf P}_\textup{S}}{\partial T}\right)_\eta + \sum_v \left( \frac{\partial{\bf P}_\textup{S}}{\partial \eta_v}\right)_T \left( \frac{d\eta_v}{dT}\right)_\sigma \nonumber \\
    &=& {\bf p}^\prime(T) + {\bf p}^{\prime\prime}(T) \; ,
\end{eqnarray}
where the primary term is evaluated at constant strain, $\eta$, and is thus due to thermally-induced atomic displacements occurring within the clamped lattice, and where the secondary effect arises from the coupled thermal expansion and piezoelectric response of the lattice:
\begin{equation}
\label{eq:second}
p^{\prime\prime}_c = e_{cv} \alpha_v \; , 
\end{equation}
where $e_{cv} = \partial P_{\textup{S},c}/\partial \eta_v$ are piezoelectric stress constants, $\alpha_v = \partial \eta_v /\partial T$ are elements of the thermal expansion tensor, $c=1,2,3$ is a Cartesian index, $v=1,\dots,6$ is a Voigt index, and where the Einstein summation convention is used. As such, the secondary term of Eq. (\ref{eq:second}) can be evaluated from first-principles calculations by combining quantities that can be obtained from well-established approaches in solid-state physics, such as the modern theory of polarization and the quasi-harmonic approximation.\cite{liu2025bond,liu2025bond2,liu2019origin,jain2023quasi,erba2013piezoelectricity,mahmoud2014low,erba2017nuclear,erba2016internal,baima2016direct,baroni2010density,allen2020theory,erba2015anharmonic,erba2015differently,erba2014combining} 

The explicit calculation of the primary term, ${\bf p}^{\prime}(T)$, from first principles proves more challenging as it calls for a treatment of anharmonicity in the phonon dynamics, as we shall discuss below. To describe how the spontaneous polarization of the clamped lattice changes as a function of atomic displacements, we expand it to first order in the phonon normal modes:\cite{liu2016first} 
\begin{equation}
\label{eq:polexp}
{\bf P}_\textup{S} \simeq {\bf P}_{\textup{S},0} + \sum_{i{\bf q}} \frac{\partial {\bf P}_\textup{S}}{\partial Q_{i{\bf q}}} Q_{i{\bf q}} \; .
\end{equation}
The primary pyroelectric coefficient thus takes the following form:
\begin{equation}
    \label{eq:pyroprim}
    {\bf p}^\prime(T) \simeq \sum_i \frac{\partial {\bf P}_\textup{S}}{\partial Q_{i{\bf 0}}}  \frac{d \langle Q_{i{\bf 0}}\rangle_T}{d T} \equiv \frac{1}{V}\sum_i \overline{{\bf Z}}_{i{\bf 0}}^\ast  \frac{d \langle Q_{i{\bf 0}}\rangle_T}{d T} \; ,
\end{equation}
where the sum runs over phonons at the Brillouin zone center $\Gamma$, $\overline{{\bf Z}}_{i{\bf 0}}^\ast$ is the Born effective charge tensor in the basis of normal mode coordinates for the $i$-th phonon at $\Gamma$ -- all phonons with ${\bf q} \neq {\bf 0}$ have vanishing $\overline{{\bf Z}}_{i{\bf q}}^\ast$ and thus do not contribute to Eq. (\ref{eq:pyroprim}) -- and $\langle \cdots \rangle_T$ denotes a thermal average, which can be expressed from statistical thermodynamics as:
\begin{equation}
    \label{eq:thermaverage}
     \langle Q_{i{\bf 0}}\rangle_T = \sum_s p(s;T) \langle Q_{i{\bf 0}}\rangle_s \quad \textup{with} \quad p(s;T) = \frac{1}{Z} e^{-\beta E_s} \; ,
\end{equation}
where the sum runs over all vibrational states from Eq. (\ref{eq:nucSchro}), $\langle Q_{i{\bf 0}}\rangle_s \equiv \langle s\vert \hat{Q}_{i{\bf 0}}\vert s \rangle$, $\beta = (k_\textup{B}T)^{-1}$ is Boltzmann's inverse temperature factor, and $Z = \sum_t e^{-\beta E_t}$ is the canonical vibrational partition function.

Let us note that, in the presence of a harmonic potential as in Eq. (\ref{PesTaylor0}), all $\langle Q_{i{\bf 0}}\rangle_s$ would vanish and so would the primary pyroelectric coefficient ${\bf p}^\prime(T)$ in Eq. (\ref{eq:pyroprim}). An explicit anharmonic description of phonon states is thus required to get a non-vanishing ${\bf p}^\prime(T)$ from Eq. (\ref{eq:pyroprim}). We shall now illustrate how the methods for anharmonic phonons presented in Section \ref{sec:anharm} can be applied for this purpose.

\subsubsection{p-VPT}
\label{sec:pyropt}

We apply the perturbation theory approach outlined in Section \ref{sec:vpt2} to the calculation of pyroelectric quantities in Eqs. (\ref{eq:pyroprim}) and (\ref{eq:thermaverage}). Let us recall that within the p-VPT approach, nuclear states are labeled by vibrational configuration vectors so that $\vert s \rangle \equiv \vert {\bf n} \rangle$ as per Eq. (\ref{eq:states}). We start by describing the expectation value of the $\hat{Q}_{i{\bf 0}}$ operator for a given state. To second order it can be written as:
\begin{eqnarray}
\label{eq:PTav0}
   \langle {\bf n}\vert \hat{Q}_{i{\bf 0}}\vert {\bf n} \rangle &\simeq&  \langle {\bf n}^{(0)}\vert \hat{Q}_{i{\bf 0}}\vert {\bf n}^{(0)} \rangle \\
   \label{eq:PTav1}
   &+& 2\lambda \langle {\bf n}^{(1)}\vert \hat{Q}_{i{\bf 0}}\vert {\bf n}^{(0)} \rangle \\
   \label{eq:PTav2}
   &+& \lambda^2 \left[ \langle {\bf n}^{(1)}\vert \hat{Q}_{i{\bf 0}}\vert {\bf n}^{(1)} \rangle + 2 \langle {\bf n}^{(2)}\vert \hat{Q}_{i{\bf 0}}\vert {\bf n}^{(0)} \rangle\right] \;\;\;\;\\
   \label{eq:PTavtot}
   &\simeq& 2\lambda \langle {\bf n}^{(1)}\vert \hat{Q}_{i{\bf 0}}\vert {\bf n}^{(0)} \rangle \; , 
\end{eqnarray}
where the zeroth-order, harmonic, term in Eq. (\ref{eq:PTav0}) vanishes, as do the two second-order terms in Eq. (\ref{eq:PTav2}). Only the first-order term of Eq. (\ref{eq:PTav1}) does not vanish -- see Appendix \ref{app:pt12} for the calculation of all terms -- yielding the following expression:
\begin{equation}
    \label{eq:qi0n}
    \braket{{\bf n}^{(1)} | \hat{Q}_{i{\bf 0}}| {\bf n}^{(0)}} 
    = - \frac{\hbar}{2} \sum_{j{\bf q}} 
    \Phi_{i{\bf 0} j\mathbf{\bf q} j-{\bf q}}
    \frac{2 n_{j{\bf q}} + 1}{4\omega_{j {\bf q}} \omega_{i {\bf 0}}^2} \; ,
\end{equation}
where the sum runs over all phonons within the first Brillouin zone (apart from the three acoustic modes at $\Gamma$), and where it can be seen that only cubic anharmonic force constants of type $\Phi_{i{\bf 0} j\textbf{{\bf q}} j-{\bf q}}$ enter the expression. We discuss the calculation of these cubic force constants in Appendix \ref{app:phi}.

We now address the perturbative expression for Boltzmann's population probability of each state, $p({\bf n};T)$. By substituting Eq. (\ref{eq:perte}) into Eq. (\ref{eq:thermaverage}) and by Taylor expansion to second order in $\lambda$, we get:
\begin{eqnarray}
  p({\bf n};T) &\simeq&  \frac{1}{Z^{(0)}} e^{-\beta E_{\bf n}^{(0)}} \nonumber \\
  \label{eq:pnpert}
  &+& \lambda^2 \frac{\beta e^{-\beta E_{\bf n}^{(0)}}}{Z^{(0)}}  \left[ \frac{\sum_{\bf{m}} E_{{\bf m}}^{(2)} e^{-\beta E_{\bf m}^{(0)}}}{Z^{(0)}} - E_{\bf n}^{(2)}\right] \; ,
\end{eqnarray}
where $Z^{(0)} = \sum_{\bf m} e^{-\beta E_{\bf m}^{(0)}}$, and where we have taken into account that the first-order perturbative correction to the energy vanishes (i.e.\textbf{,} $E_{\bf n}^{(1)} = 0$). By combining Eq. (\ref{eq:PTavtot}) with Eq. (\ref{eq:pnpert}), grouping same-order terms, and setting $\lambda = 1$, we get the following perturbative expression for the thermal average $\langle Q_{i{\bf 0}}\rangle_T $ of Eq. (\ref{eq:thermaverage}) up to second order:
\begin{eqnarray}
\label{eq:av2}
\langle Q_{i{\bf 0}}\rangle_T &\simeq& \frac{2}{Z^{(0)}} \sum_{\bf n}   \langle {\bf n}^{(1)}\vert \hat{Q}_{i{\bf 0}}\vert {\bf n}^{(0)} \rangle  e^{-\beta E_{\bf n}^{(0)}} \\
\label{eq:avfin}
&=& - \frac{1}{N_q}\sum_{j{\bf q}} 
    \Phi_{i{\bf 0} j\textbf{{\bf q}} j-{\bf q}} \frac{\hbar}{4\omega_{j {\bf q}} \omega_{i {\bf 0}}^2}
    \frac{1 + e^{-\beta \hbar\omega_{j {\bf q}}}}{1 - e^{-\beta \hbar\omega_{j {\bf q}}}} \; ,
\end{eqnarray}
where $N_q$ is the number of sampled \textbf{${\bf q}$} points and where in the second equality we have used Eq. (\ref{eq:qi0n}) and exploited the harmonic character of the energies that enter Eq. (\ref{eq:av2}), as illustrated in Appendix \ref{app:pt12}. By taking the analytical derivative with respect to temperature $T$ of Eq. (\ref{eq:avfin}) and substituting the resulting expression into Eq. (\ref{eq:pyroprim}), we get the following final expression for the primary pyroelectric coefficient from perturbation theory up to second order:
\begin{widetext}
\begin{equation}
    \label{eq:pyroprimfinal}
    {\bf p}^\prime(T) \simeq -\frac{\hbar^2}{2k_\textup{B}T^2V}\sum_i   \frac{\overline{{\bf Z}}_{i{\bf 0}}^\ast}{\omega_{i{\bf 0}}^2}\frac{1}{N_q}\sum_{j{\bf q}} 
    \Phi_{i{\bf 0} j\textbf{{\bf q}} j-{\bf q}} \frac{e^{-\beta \hbar\omega_{j {\bf q}}}}{(1-e^{-\beta \hbar\omega_{j {\bf q}}})^2} .
\end{equation}
\end{widetext}
Let us note that the expression above is intrinsically first-order in the perturbation as all zeroth- and second-order terms vanish.

\subsubsection{VSCF}
\label{sec:vscf222}

As discussed in Section \ref{sec:vscftheo}, within the VSCF approach, nuclear states can be labeled by vibrational configuration vectors so that $\vert s \rangle \equiv \vert {\bf n} \rangle$. The expectation value of the $\hat{Q}_{i{\bf 0}}$ operator over state $\vert {\bf n} \rangle$ is obtained from:
\begin{eqnarray}
\langle {\bf n}\vert \hat{Q}_{i{\bf 0}}\vert {\bf n} \rangle &=& \langle n_{i{\bf 0}}\vert \hat{Q}_{i{\bf 0}}\vert n_{i{\bf 0}} \rangle \prod_{j{\bf q} \neq i{\bf 0}} \langle n_{j{\bf q}}\vert n_{j{\bf q}} \rangle
= \langle n_{i{\bf 0}}\vert \hat{Q}_{i{\bf 0}}\vert n_{i{\bf 0}} \rangle \nonumber \\
&=& \sum_{m_{i{\bf 0}} = 0}^{M}\sum_{l_{i{\bf 0}} = 0}^{M} C_{m_{i{\bf 0}},n_{i{\bf 0}}} C_{l_{i{\bf 0}},n_{i{\bf 0}}} \langle l_{i{\bf 0}}^{(0)}\vert \hat{Q}_{i{\bf 0}}\vert m_{i{\bf 0}}^{(0)} \rangle \, ,
\end{eqnarray}
where we have used Eqs. (\ref{eq:states}) and (\ref{eq:linVSCF}) along with the orthonormality of the one-mode wavefunctions. The thermal average of the normal coordinate $\hat{Q}_{i{\bf 0}}$ is given by:
\begin{equation}
\label{eq:ggg}
   \langle Q_{i{\bf 0}}\rangle_T = \frac{1}{Z} \sum_{\bf n} \langle {\bf n}\vert \hat{Q}_{i{\bf 0}}\vert {\bf n} \rangle e^{-\beta E_{\bf n}} \; , 
\end{equation}
where $Z = \sum_{\bf m} e^{-\beta E_{\bf m}}$ and the energy of each configuration is given by Eq. (\ref{eq:vscfene}). In contrast to Section \ref{sec:pyropt} for the p-VPT approach, where the energies of the states entering the thermal average of Eq. (\ref{eq:av2}) were still harmonic and thus the infinite sum over configurations could take a simple closed analytical form (as discussed in Appendix \ref{app:pt12}), here the vibrational configurational space needs to be sampled more explicitly.
Although possible in principle, it proves impractical to converge $Z$ by explicit summation over enumerated configurations. We instead sample the configurational space by use of a stochastic Metropolis-Hastings approach\cite{metropolis1953equation,hastings1970monte} that we illustrate in Appendix \ref{app:MH}. For a given temperature, the algorithm selects a sequence of $\tilde{N}$ vibrational configurations $\tilde{\bf n}$ that are representative of the underlying -- and yet unknown in its normalization factor -- Boltzmann probability distribution such that the thermal average in Eq. (\ref{eq:ggg}) becomes:
\begin{equation}
\label{eq:uuu}
   \langle Q_{i{\bf 0}}\rangle_T \simeq \frac{1}{\tilde{N}} \sum_{\tilde{\bf n}} \langle \tilde{\bf n}\vert \hat{Q}_{i{\bf 0}}\vert \tilde{\bf n} \rangle  \; . 
\end{equation}
Finally, to compute the pyroelectric coefficient from Eq. (\ref{eq:pyroprim}), we evaluate the derivative with respect to temperature of Eq. (\ref{eq:uuu}) numerically.

\subsubsection{VCI}
\label{sec:pyrovci}

As illustrated in Section \ref{sec:vscftheo}, within the VCI approach each vibrational state $s$ is expressed as a linear combination of configuration states, see Eq. (\ref{eq:vcifunc}). Within the VCI-HO variant of the method, the expectation value of the $\hat{Q}_{i{\bf 0}}$ operator over a state $\vert s \rangle$ can be written as:
\begin{widetext}
\begin{eqnarray}
\langle s\vert \hat{Q}_{i{\bf 0}}\vert s \rangle &=& \sum_{{\bf n}} \sum_{{\bf m}} A_{{\bf n},s} A_{{\bf m},s} \langle {\bf m}\vert \hat{Q}_{i{\bf 0}}\vert {\bf n}  \rangle \nonumber \\
&=& \sum_{{\bf n}} \sum_{{\bf m}} A_{{\bf n},s} A_{{\bf m},s}  \langle {m_{i{\bf 0}}}^{(0)}\vert \hat{Q}_{i{\bf 0}}\vert {n_{i{\bf 0}}}^{(0)}  \rangle \prod_{j{\bf q}}\!\left.^{\prime}\right. \langle {m_{j{\bf q}}}^{(0)}\vert {n_{j{\bf q}}}^{(0)}  \rangle \nonumber \\
&=& \sum_{{\bf n}} A_{{\bf n},s} \left[A_{{\bf n} + {\bf 1}_{i{\bf 0}},s} \langle ({n_{i{\bf 0}}+1)^{(0)}}\vert \hat{Q}_{i{\bf 0}}\vert {n_{i{\bf 0}}}^{(0)}  \rangle + A_{{\bf n} - {\bf 1}_{i{\bf 0}},s} \langle ({n_{i{\bf 0}}-1)^{(0)}}\vert \hat{Q}_{i{\bf 0}}\vert {n_{i{\bf 0}}}^{(0)}  \rangle\right] \nonumber \\
\label{eq:www}
&=& \sum_{{\bf n}} A_{{\bf n},s} \left[A_{{\bf n} + {\bf 1}_{i{\bf 0}},s}\sqrt{\hbar\frac{n_{i{\bf 0}} + 1}{2 \omega_{i{\bf 0}}}} + A_{{\bf n} - {\bf 1}_{i{\bf 0}},s}  \sqrt{\hbar\frac{n_{i{\bf 0}}}{2 \omega_{i{\bf 0}}}} \,\right] \; ,
\end{eqnarray}
\end{widetext}
where the primed product runs over all phonons with $j{\bf q} \neq i{\bf 0}$, and ${\bf 1}_{i{\bf 0}} = (0, \dots, 0, 1_{i{\bf 0}}, 0, \dots, 0)$ is a unit configuration vector with one quantum of excitation in mode $i{\bf 0}$ and zero in all other modes. The derivation strategy followed in Eq. (\ref{eq:www}) is similar to that discussed in Appendix \ref{app:pt12} for Eq. (\ref{eq:ppp}).

The computational cost of VCI (i.e.\textbf{,} building and diagonalizing the nuclear Hamiltonian matrix \textbf{${\bf H}$} in the basis of the vibrational configurations) grows combinatorially with the number of vibrational degrees of freedom, which prevents its use for practical calculations on real systems. However, VCI can be useful on toy systems where it can provide exact reference solutions. In particular, we shall use it in Section \ref{sec:resu} to compute the pyroelectric coefficient of ZnO from a single-phonon approximation in order to compare key aspects of perturbative versus variational methods. In this case, the partition function $Z$ and thermal average $\langle Q_{i{\bf 0}}\rangle_T$ from Eq. (\ref{eq:thermaverage}) can be converged by direct summation over states.

\section{Computational Details}
\label{sec:compdet}

All calculations are performed with a developmental version of \textsc{Crystal23},\cite{Erba2023} a package for electronic structure theory of materials that makes use of a basis set of local, atom-centered, Gaussian-type functions within a periodic linear combination of atomic orbitals (LCAO) approach. To benchmark the methodology presented in Section \ref{sec:theory}, calculations are performed on ZnO in its wurtzite structure (space group \(P6_{3}mc\)). The dependence of the results on the treatment of exchange and correlation is assessed by employing three functionals, each representative of a different rung of Jacob's ladder: the local density approximation (Slater exchange with Vosko-Wilk-Nusair correlation),\cite{Kohn1965, Vosko1980} the generalized gradient approximation by Perdew-Burke-Ernzerhof, PBE,\cite{pbe_art} and its global hybrid counterpart PBE0, which includes 25\% exact exchange.\cite{pbe0_art} A Gaussian-type triple-$\zeta$-valence plus polarization (TZVP) basis set,\cite{kuklin2022structural} derived from Karlsruhe def2 bases,\cite{weigend2005balanced} is consistently used for all calculations; see Supplemental Material (SM).\cite{ESIPYRO} The reciprocal space is sampled using a $12 \times 12 \times 12$ Monkhorst-Pack mesh, yielding 133 ${\bf k}$ points in the symmetry-irreducible portion of the first Brillouin zone, for primitive cell calculations (see SM for {\bf k}-meshes used for supercell calculations).
Convergence on energy of the self-consistent field procedure, DFT integration grids and optimization convergence thresholds were set to default values for all calculations. The thresholds for the truncation of Coulomb and exchange infinite series are reported in the \textsc{Crystal} input deck available in the SM.  
Harmonic frequencies were computed numerically via a central finite-difference approach with an atomic displacement of 0.001 \AA. Born effective charge tensor components are computed through density functional perturbation theory (DFPT).\cite{maschio2012ab}

\section{Results and Discussion}
\label{sec:resu}

We perform calculations on ZnO in its thermodynamically stable wurtzite phase\textbf{,} which crystallizes in a hexagonal lattice belonging to the space group \(P6_{3}mc\).\cite{geurts2010crystal} The crystal structure consists of two interpenetrating hexagonal close-packed (hcp) sublattices of nominal \(\mathrm{Zn}^{2+}\) cations and \(\mathrm{O}^{2-}\) anions, mutually displaced along the {\bf c} axis by the fractional internal parameter \(u\) with an ``ideal'' value of $\frac{3}{8} = 0.375$. Each chemical species occupies local sites of \(C_{3v}\) point symmetry, forming a network of edge-sharing \(\mathrm{ZnO}_{4}\) tetrahedra.\cite{ozgur2005comprehensive} A key physical consequence of the \(P6_{3}mc\) space group symmetry is the absence of an inversion center along the {\bf c} axis and simultaneously of a mirror plane or $C_2$ axis perpendicular to it, which imparts a structural deviation of the internal parameter \(u\) from its ideal geometry and induces a spontaneous macroscopic polarization along the same axis.\cite{geurts2010crystal,ozgur2005comprehensive,bernardini1997spontaneous,oba2008defect} ZnO has become a benchmark for the theoretical modeling of \textbf{} polarization in solids.\cite{dal1994ab,massidda1995polarization,hill2000first,catti2003full,wu2005systematic, liu2018internal,masuki2023full,rostami2025anisotropic,gong2026ab}

\begin{figure}[t!]
    \centering
    \includegraphics[width=1.0\linewidth]{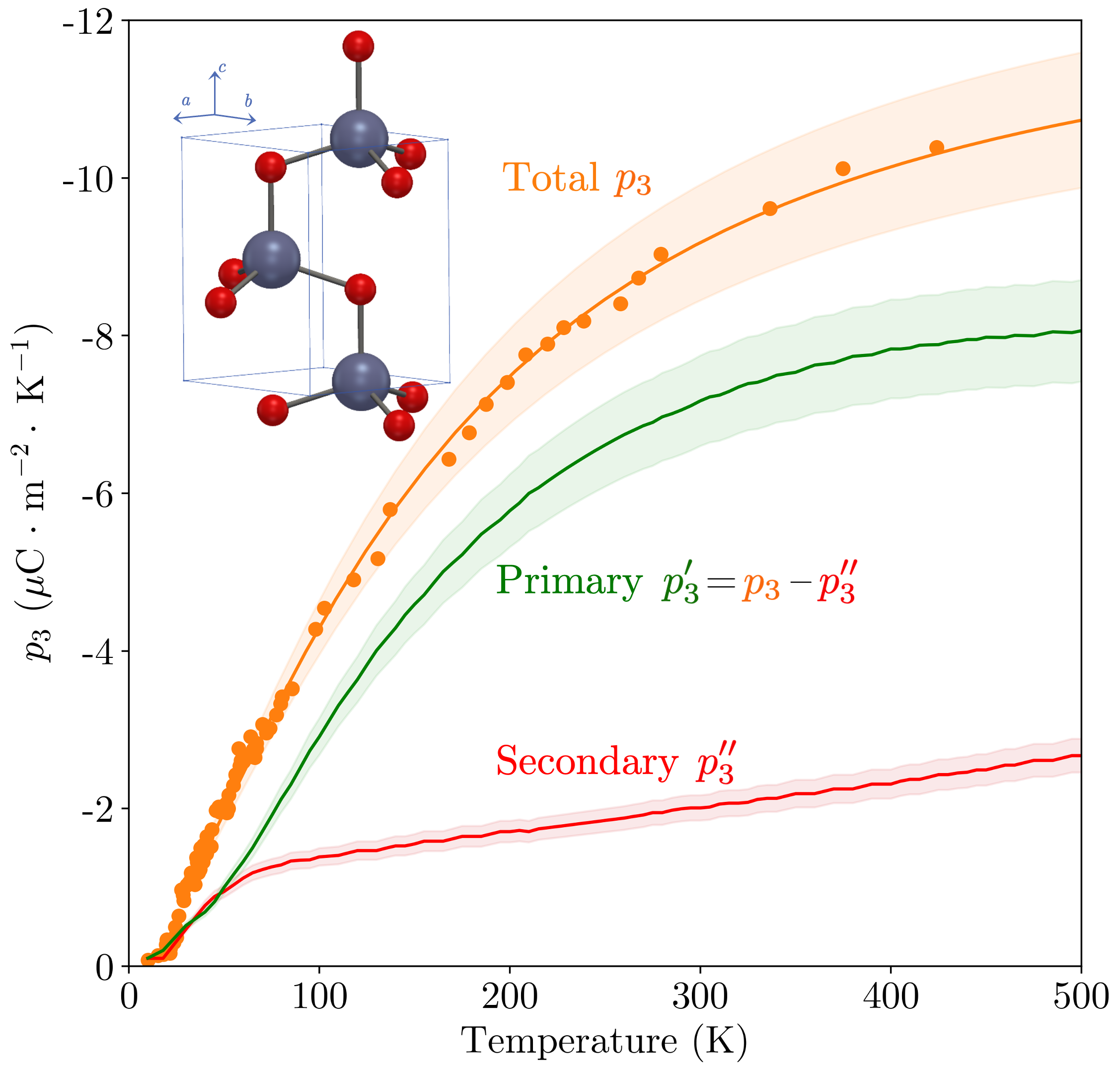}
    \caption{Temperature dependence of the pyroelectric coefficient of bulk ZnO. Experimental data by Heiland and Ibach~\cite{heiland1966pyroelectricity} for the total pyroelectric coefficient, $p_3(T)$, \textbf{are} shown in orange. The secondary coefficient, $p_3^{\prime\prime}(T)$, is reported in red as recently estimated by Gong and Dal Corso~\cite{gong2026ab} from consolidated piezoelectric and thermal expansion data. The primary coefficient, obtained by taking the difference between the total and secondary effects, $p^{\prime}_3(T) =p_3(T) - p^{\prime\prime}_3(T)$, is reported in green. Shaded regions represent a nominal uncertainty of \textbf{$\pm 8\%$}. The inset shows the crystal structure of wurtzite-type ZnO (space group \(P6_{3}mc\)).}
    \label{fig:pyrexp}
\end{figure}

Pyroelectricity in ZnO consistently occurs along the {\bf c} axis, with a total, constant-stress pyroelectric coefficient, $p_3(T)$, that has been measured experimentally by Heiland and Ibach in the temperature range 9.7--420 K,\cite{heiland1966pyroelectricity} taking a value of \textbf{$-9.4 \pm 0.7$} $\mu \textup{C} \cdot \textup{m}^{-2} \cdot \textup{K}^{-1}$ at room temperature. The primary, constant-strain effect cannot be accessed directly from available experiments and instead needs to be evaluated from the difference between the total and the secondary effects, $p^{\prime}_3(T) =p_3(T) - p^{\prime\prime}_3(T)$, where the secondary contribution can be effectively derived from the measured piezoelectric stress tensor {\bf e} and the anisotropic thermal expansion tensor ${\boldsymbol \alpha}$, as per Eq. (\ref{eq:second}): $p^{\prime\prime}_3(T) = 2e_{31}(T)\alpha_{1}(T) + e_{33}(T)\alpha_3(T)$. A first attempt at estimating the secondary effect $p^{\prime\prime}_3(T)$ resulted in the large value of \textbf{$-5.2$} $\mu \textup{C} \cdot \textup{m}^{-2} \cdot \textup{K}^{-1}$ at room temperature.\cite{albertsson1989atomic} Recently, more reliable estimates based on consolidated data for {\bf e} and ${\boldsymbol \alpha}$ have been reported with a smaller value of \textbf{$-2.0 \pm 0.2$} $\mu \textup{C} \cdot \textup{m}^{-2} \cdot \textup{K}^{-1}$ at room temperature.\cite{liu2016first,masuki2023full,gong2026ab} This leads to an estimated value for the primary coefficient $p_3^\prime(T)$ of approximately \textbf{$-7.0 \pm 0.6$} $\mu \textup{C} \cdot \textup{m}^{-2} \cdot \textup{K}^{-1}$ at room temperature. We show the temperature dependence of the primary, secondary, and total pyroelectric coefficients of ZnO in Figure \ref{fig:pyrexp}. The secondary coefficient shows an almost linear trend with temperature above 100 K.\cite{liu2016first,masuki2023full,gong2026ab} The primary coefficient exhibits the same temperature dependence as the constant-volume specific heat: it approaches zero as $T^3$ and saturates to a constant value at high temperature.\cite{heiland1966pyroelectricity,yan2009contributions}   

\begin{table}[b!]
\caption{Selected structural and dynamical parameters of wurtzite ZnO: the lattice parameter $a$ (in \AA), the $c/a$ ratio, the internal parameter $u$, the perpendicular, $Z^\ast_\perp$, and parallel, $Z^\ast_\parallel$, Born effective charges (in $e$), and the A$_1$ (TO) phonon frequency at $\Gamma$ (in cm$^{-1}$). Present results from DFT calculations with different xc functionals are compared to available experimental data.}
\label{tab:initial}
\begin{tabular}{lllllll}
\hline\hline
&&&&&&\\
Method & \multicolumn{1}{c}{$a$} & \multicolumn{1}{c}{$c/a$} & \multicolumn{1}{c}{$u$} & \multicolumn{1}{c}{$Z^\ast_\perp$} & \multicolumn{1}{c}{$Z^\ast_\parallel$} & $\omega_{\textup{A}_1\textup{ (TO)}}$ \\
&&&&&&\\
LDA  & 3.205 & 1.604 & 0.3808 & 2.02 & 2.08 & 418\\
PBE  & 3.294 & 1.600 & 0.3818 & 2.05 & 2.11 & 372\\
PBE0 & 3.266 & 1.596 & 0.3824 & 2.01 & 2.07 & 397\\
&&&&&&\\
Exp. &3.246~\cite{yoshio2001crystal}&1.603~\cite{yoshio2001crystal}&0.3812~\cite{yoshio2001crystal}&2.02~\cite{PhysRevB.65.092101}&2.10~\cite{madelung1982numerical,PhysRevB.99.104101} &380~\cite{ozgur2005comprehensive}\\
&&&&&&\\
\hline\hline  
\end{tabular}
\end{table}

\subsection{Initial Structural and Dynamical (Harmonic) Assessment of ZnO}
\label{sec:prelim}

Full structural relaxations of wurtzite-type ZnO are performed in terms of both atomic positions and lattice cell parameters,\cite{civalleri2001hartree,erba2016internal} with the three xc functionals here considered: LDA, PBE, and PBE0. The Born effective charge tensor is computed from DFPT.\cite{maschio2012ab} Harmonic phonon frequencies are computed from the FP approach\cite{PyrGro} outlined in Section \ref{sec:harm}. The computed values of selected structural and dynamical parameters of wurtzite ZnO obtained using different xc functionals are summarized in Table \ref{tab:initial}, alongside available low-temperature experimental benchmarks. 
As expected, the LDA exhibits its characteristic overbinding tendency, underpredicting the lattice parameter \(a\) (\(3.205 \text{ \AA}\) versus the experimental value of \(3.246 \text{ \AA}\)) and consequently overestimating the \(\text{A}_1\text{(TO)}\) phonon frequency (\(418\text{ cm}^{-1}\) compared to the experimental \(380\text{ cm}^{-1}\)). Conversely, the GGA (in its PBE parametrization) overcorrects this trend, underbinding the lattice (\(a = 3.294\text{ \AA}\)) and slightly underestimating the \(\text{A}_1\text{(TO)}\) mode frequency (\(372\text{ cm}^{-1}\)). The hybrid functional PBE0 mitigates this tendency, yielding an optimized lattice constant of \(3.266\text{ \AA}\) in closer alignment with experiment. Regarding the dielectric properties, the magnitude and anisotropy of the Born effective charges (\(Z^*_{\parallel} > Z^*_{\perp}\)) are robustly captured across all functionals, demonstrating good quantitative agreement with the experimental data.

\begin{figure}[t!]
    \centering
    \includegraphics[width=1.0\linewidth]{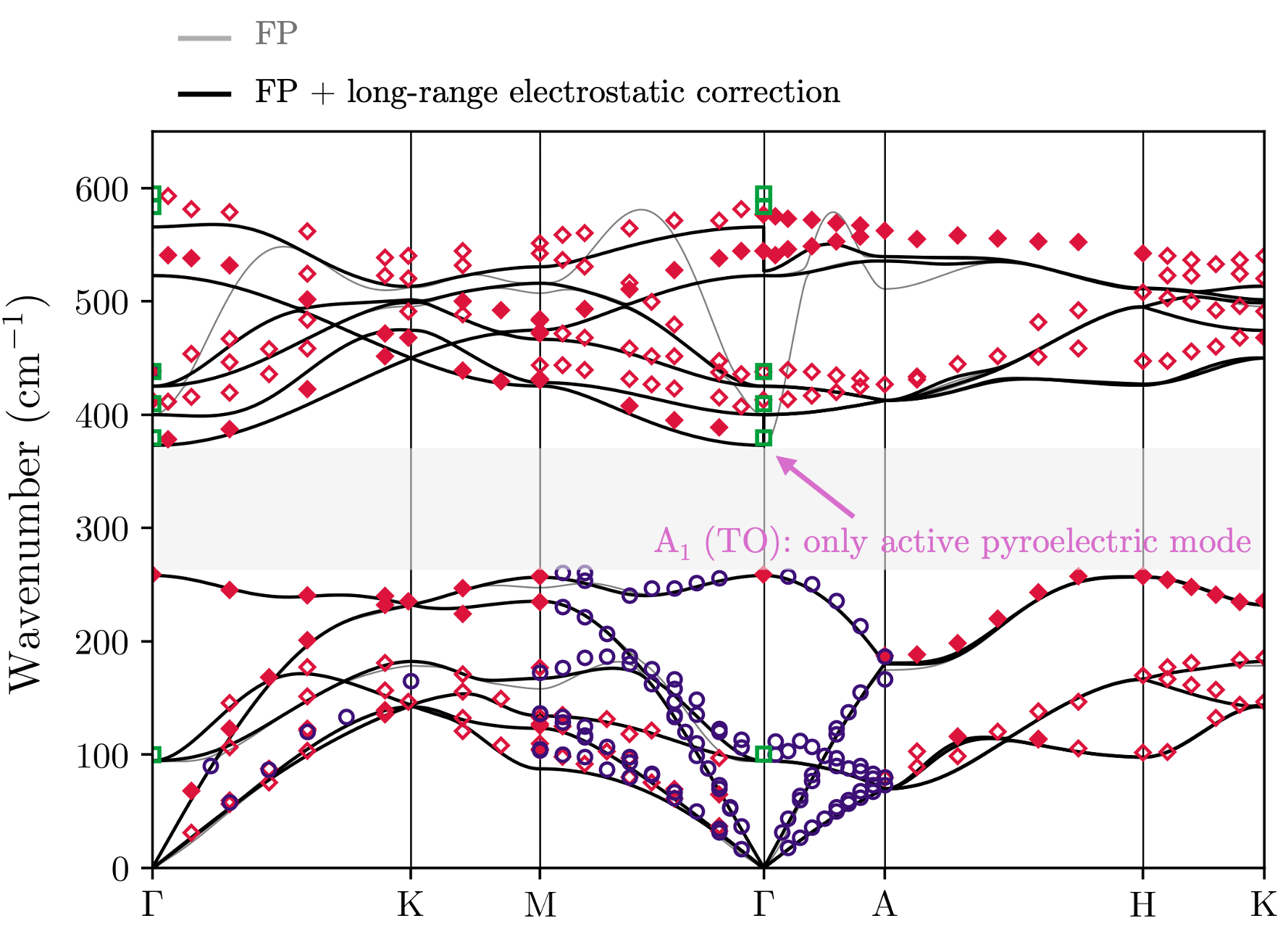}
    \caption{Phonon band structure of bulk ZnO in its wurtzite structure. The gray and black solid lines are the harmonic phonon band structures obtained from present frozen phonon (FP) calculations by interpolation from a $5\times 5\times 5$ mesh without and with the correction for long-range electrostatic interactions described in Section \ref{sec:long}, respectively. Results are shown for the PBE functional. Available experimental data are reported as symbols: red diamonds from inelastic neutron scattering at 10 K,~\cite{serrano2010phonon} blue circles from inelastic neutron scattering,~\cite{hewat1970lattice,thoma1974lattice} and green squares from Raman scattering at 7 K.\cite{serrano2003dependence} This figure was produced with the help of the \textsc{Crystal}Clear Python interface to \textsc{Crystal}.\cite{CRYSTALpytools}}
    \label{fig:bands}
\end{figure}

For each considered xc functional, we have computed harmonic phonons with the FP approach described in Section \ref{sec:harm} on a $5\times 5\times 5$ supercell of the primitive lattice (500 atoms in the supercell), which ensures a converged description of the phonon dispersion, as shown in the SM.\cite{ESIPYRO} We show the computed PBE harmonic phonon band structure in Figure \ref{fig:bands} as obtained from the FP approach without (gray solid line) and with (black solid line) the ``non-analytical'' correction for long-range electrostatic interactions described in Section \ref{sec:long}. Available experimental data from low-temperature inelastic neutron scattering and Raman scattering are reported as symbols.~\cite{serrano2010phonon,hewat1970lattice,thoma1974lattice,serrano2003dependence} Several key features emerge: i) acoustic branches, and in general modes below the gap, are described very accurately; ii) the long-range electrostatic correction is key to a correct description of certain bands as they approach $\Gamma$, and of the LO-TO splitting; iii) the shape of all optical bands is consistent with the experiments; iv) phonon frequencies from the top optical bands are slightly underestimated by PBE. A comparison among the LDA, PBE, and PBE0 harmonic phonon band structures is provided in the SM.

As per Eq. (\ref{eq:pyroprim}), in wurtzite-structured ZnO there is a single phonon mode of A$_1$ symmetry at $\Gamma$ with a non-vanishing contribution to the primary pyroelectric coefficient, with a phonon frequency of about 380 cm$^{-1}$ (in its transverse optical, TO, branch),\cite{ozgur2005comprehensive} see the pink arrow in Figure \ref{fig:bands}. We shall refer to this mode as the pyroelectrically active mode. As far as phonon couplings with the pyroelectrically active mode are considered, acoustic modes are expected to dominate the pyroelectric coefficient at low temperature, while optical modes to contribute progressively more as temperature increases.\cite{yan2009contributions,liu2016first} Optical modes have been estimated to account for 25\% and 33\% of the primary pyroelectric coefficient of wurtzite-structured ZnO\cite{liu2016first} and GaN,\cite{yan2009contributions} respectively, at room temperature.

\subsection{The Single Phonon Approximation to Primary Pyroelectricity}
\label{sec:BPA}

As discussed at the end of Section \ref{sec:prelim}, there is a single pyroelectrically active mode of $A_1$ symmetry at $\Gamma$ in wurtzite-structured ZnO, and thus Eq. (\ref{eq:pyroprim}) reduces to:
$ {\bf p}^\prime(T) \simeq \frac{1}{V} \overline{{\bf Z}}_{\textup{A}_1}^\ast  \frac{d \langle Q_{\textup{A}_1}\rangle_T}{d T}$. As such, one may be tempted by a single-phonon description of primary pyroelectricity in ZnO. Let us explore this strategy -- which will prove rather informative -- where only the phonon anharmonicity of the A$_1$ (TO) mode is considered. 
All other phonons are assumed to be harmonic and all couplings among phonons are assumed to be null (i.e., phonons are assumed to be independent). This reduces the problem to a one-dimensional toy system, whose exact solution can be targeted with the VCI approach~\cite{NOTEVCIVSCF} described in Section \ref{sec:pyrovci} and used to rigorously assess the p-VPT perturbative approach of Section \ref{sec:pyropt}. Within the single phonon approximation (SPA), the expression for the p-VPT primary pyroelectric coefficient of Eq. (\ref{eq:pyroprimfinal}) in ZnO is simplified to:
\begin{equation}
    \label{eq:pyroprimfinalBPA}
    {\bf p}^\prime_\textup{SPA}(T) \simeq -\frac{\hbar^2}{V} \overline{{\bf Z}}_{\textup{A}_1}^\ast  \frac{ \Phi_{\textup{A}_1 \textup{A}_1 \textup{A}_1}}{2k_\textup{B}T^2\omega_{\textup{A}_1}^2} \frac{e^{-\beta \hbar\omega_{\textup{A}_1}}}{(1-e^{-\beta \hbar\omega_{\textup{A}_1}})^2} \; , 
\end{equation}
where $\Phi_{\textup{A}_1 \textup{A}_1 \textup{A}_1}$ is the cubic force constant of the A$_1$ (TO) phonon at $\Gamma$, which can be easily computed from finite-difference expressions, for instance.\cite{PARTI_ANHARM} Formally, the SPA consists in including just the self term (ST) while neglecting the mixed terms (MT) in the expression of $\langle Q_{\textup{A}_1}\rangle_T$ derived in Appendix \ref{app:pt12}. Let us stress that Eq. (\ref{eq:pyroprimfinalBPA}) just requires a $\Gamma$-only calculation and is thus characterized by a rather low associated computational cost.

\begin{figure}[t!]
    \centering
    \includegraphics[width=1.0\linewidth]{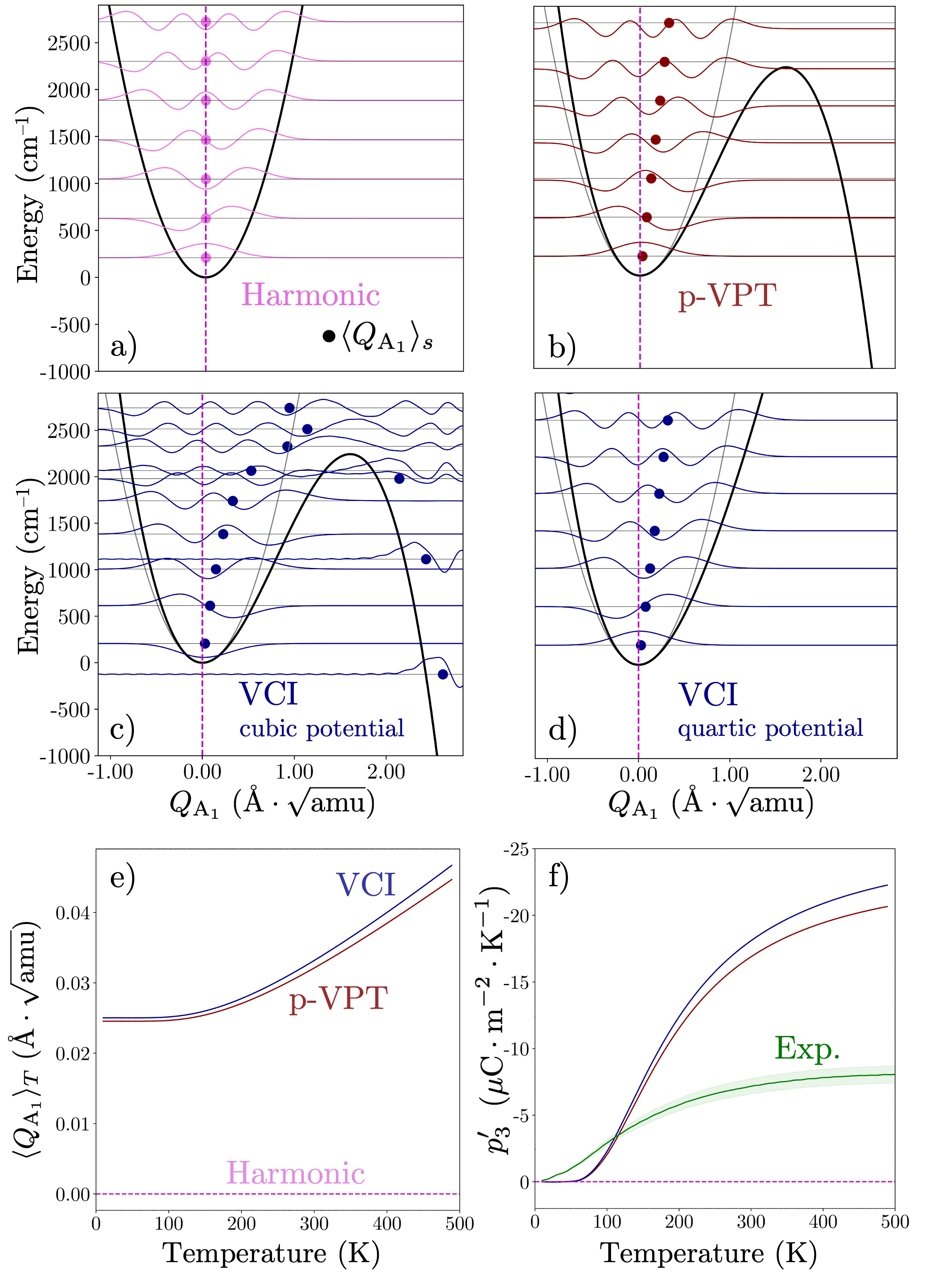}
    \caption{Single-phonon description of primary pyroelectricity in ZnO. a) The harmonic potential of the A$_1$ (TO) mode, along with the corresponding harmonic energy levels and wavefunctions for the lowest-lying vibrational states. Full circles mark the mean normal coordinate for each state, $\langle Q_{\textup{A}_1}\rangle_s$. b) As in a) from the perturbative approach, p-VPT, of Section \ref{sec:pyropt}. c) As in a) from the variational VCI method of Section \ref{sec:pyrovci} for a cubic potential. d) As in c) from VCI for a quartic potential. e) Temperature dependence of the finite-temperature normal coordinate, $\langle Q_{\textup{A}_1}\rangle_T$. f) Temperature dependence of the primary pyroelectric coefficient $p_3^\prime(T)$; the green line represents the target values from Figure \ref{fig:pyrexp}. In panels e) and f), the VCI values refer to the quartic potential of panel d). Results are shown for the LDA. In panels \textbf{b)--d)} the harmonic potential is shown as a light gray line.
    This figure was produced with the help of the \textsc{CrystalClear} Python interface to \textsc{Crystal}.\cite{CRYSTALpytools}
    }
    \label{fig:single}
\end{figure}

Let us discuss the SPA with the help of Figure \ref{fig:single}. The one-dimensional harmonic potential of the A$_1$ (TO) phonon of ZnO is shown in Figure \ref{fig:single}a) as computed within the LDA, along with the corresponding harmonic energy levels and harmonic vibrational wavefunctions for the lowest-lying states. The mean normal coordinate, $\langle Q_{\textup{A}_1}\rangle_s$, for each state $s$ is marked by a circle. For the symmetric harmonic potential, the mean is trivially always zero. An analogous representation is shown in Figure \ref{fig:single}b) for the perturbative description, where the cubic term of the potential is included, see Eq. (\ref{eq:pyroprimfinalBPA}). As per the p-VPT theory of pyroelectricity of Section \ref{sec:pyropt}, the energies of the states are still harmonic while the wavefunctions are corrected to first order from Eq. (\ref{eq:n1}). The same representation is reported in Figures \ref{fig:single}c) and \ref{fig:single}d) for the variational VCI method for a cubic and quartic representation of the potential, respectively. In both cases, states are described as a linear combination of 70 harmonic basis functions in Eq. (\ref{eq:vcifunc}). Quite clearly, even though only odd terms of the potential are expected to contribute to $\langle Q_{\textup{A}_1}\rangle_s$, the description differs in the two cases. Indeed, because of the variational nature of the VCI approach, the purely cubic potential of Figure \ref{fig:single}c) may result in unphysical solutions because of its unbounded character. In VSCF, it has been documented that unbounded mean-field potentials may even cause non-convergence.~\cite{yang2023pitfalls,spencer2026converging}
Comparison of Figures \ref{fig:single}b) and \ref{fig:single}d) shows that the description of the mean normal coordinate, $\langle Q_{\textup{A}_1}\rangle_s$, for the different states from the p-VPT and VCI approaches is extremely consistent. Panels e) and f) of Figure \ref{fig:single} show the temperature dependence of the finite-temperature normal coordinate, $\langle Q_{\textup{A}_1}\rangle_T$, and primary pyroelectric coefficient, $p^\prime_3(T)$, respectively, as computed from the reference VCI approach (blue line) and the perturbative p-VPT approach (red line). The perturbative description closely matches the reference variational one in the whole explored temperature range, slightly underestimating both quantities at high temperature (by 4\% and 8\% at 500 K for $\langle Q_{\textup{A}_1}\rangle_T$ and $p^\prime_3(T)$, respectively). In Figure \ref{fig:single}f), we also report as a green solid line the target value for the primary pyroelectric coefficient in ZnO deduced from experimental data, as discussed with the help of Figure \ref{fig:pyrexp}. 

The SPA is found to provide a qualitatively correct description of the primary pyroelectric coefficient in the explored temperature range, with an overall underestimation at low temperature and overestimation at high temperature. The low-temperature behavior is consistent with the omission of couplings with acoustic modes as discussed at the end of Section \ref{sec:prelim}. For instance, at room temperature, the primary pyroelectric coefficient obtained from the SPA is larger than the target experimental value by a factor of just slightly above 2. All data in Figure \ref{fig:single} are obtained with the LDA; very similar trends are obtained with other xc functionals, as shown in the SM.

Let us conclude this section by highlighting two key aspects: i) Although ``even'' terms of the potential do not directly contribute to mean normal coordinates and thus to the pyroelectric response (only ``odd'' terms do), they are still required in variational treatments to avoid either unbounded total (for VCI) or unbounded mean-field (for VSCF) potentials. On the contrary, the p-VPT perturbative approach does not require them, which represents a significant computational simplification; ii) The SPA provides a qualitatively correct description of primary pyroelectricity in a wide temperature range, with quantitative deviations of the order of a factor of 2--3. We shall analyze the effect of phonon dressing\cite{fugallo2018infrared} in the next section and show how the quantitative agreement can be significantly improved. At the same time, we believe that the very low computational cost and complexity of the SPA make it a very promising approach for fast high-throughput screening in materials design for pyroelectric applications.

\subsection{The Effect of Phonon Dressing on Primary Pyroelectricity}
\label{sec:dress}

\begin{figure}[t!]
    \centering
    \includegraphics[width=1.0\linewidth]{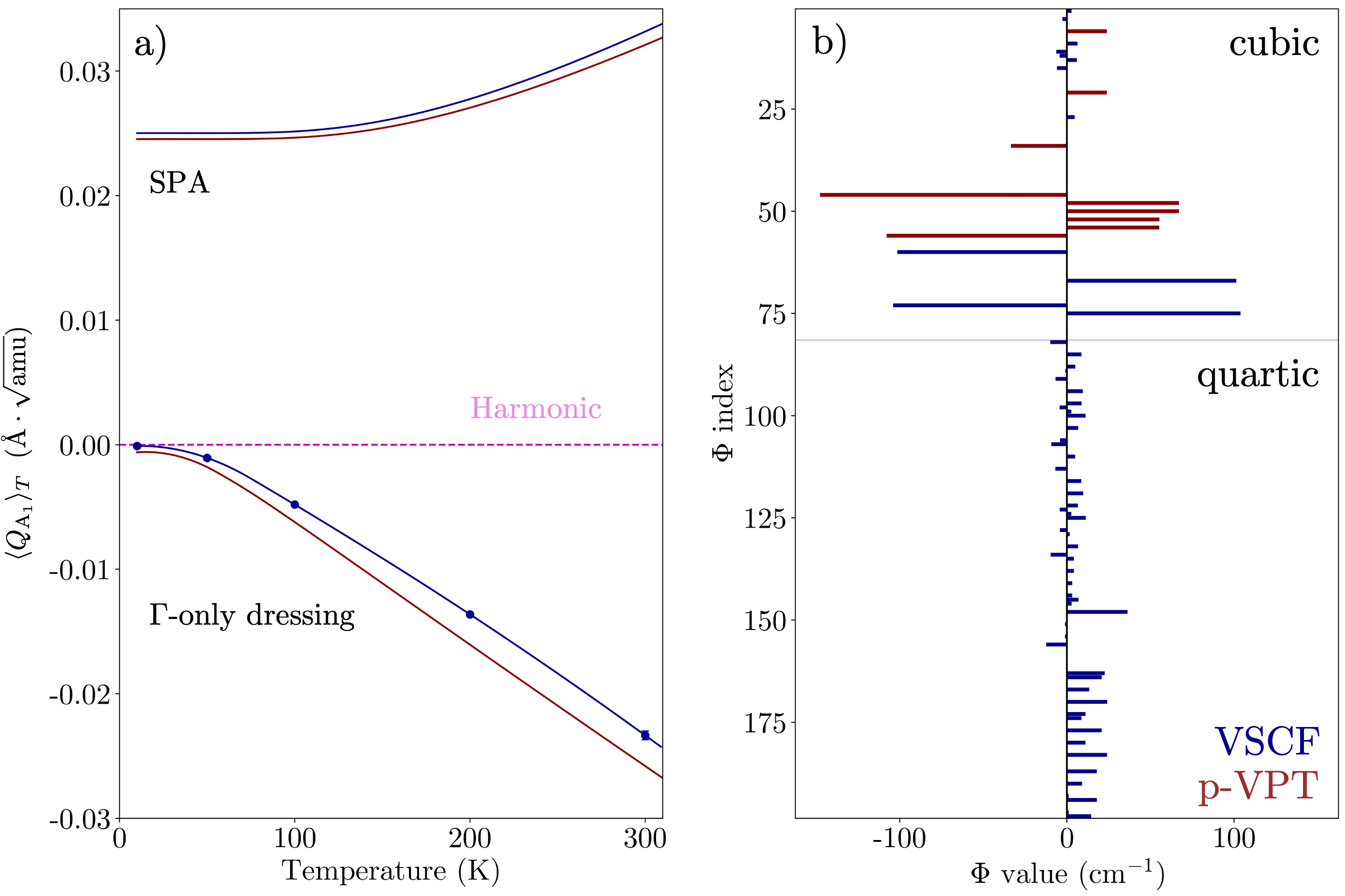}
    \caption{The effect of phonon-phonon couplings at $\Gamma$. a) Finite-temperature mean normal mode coordinate of the A$_1$ mode of ZnO by the SPA and by accounting for couplings among phonons at $\Gamma$, from the perturbative, p-VPT, approach (red lines) and the variational, VSCF, approach (blue lines). For $\Gamma$-dressed VSCF, the circles mark average values obtained for each temperature from 20 runs of the stochastic Metropolis-Hastings (MH) algorithm described in Appendix \ref{app:MH} over $\tilde{N} =$ 5 million vibrational configurations each. The spread of computed values is reported in terms of their standard deviation as vertical error bars (not visible as in all cases they are within the size of the marker). b) Sets of anharmonic force constants $\Phi$ of ZnO for phonons at $\Gamma$ used by the VSCF calculation (red and blue bars) and p-VPT calculation (red bars): they amount to 198 independent components for VSCF (81 cubic and 117 quartic) and to just 9 (cubic) for p-VPT. All values from the LDA.}
    \label{fig:Gamma}
\end{figure}

In Section \ref{sec:BPA} we introduced the single phonon approximation to primary pyroelectricity. We now analyze the effect of phonon ``dressing''~\cite{fugallo2018infrared} through the progressive inclusion of phonon-phonon couplings. We start by restricting ${\bf q} = -{\bf q} = {\bf 0}$ (i.e., by only accounting for couplings among phonon modes at $\Gamma$). The effect is shown in Figure \ref{fig:Gamma}a) on the temperature dependence of the mean normal coordinate $\langle Q_{\textup{A}_1}\rangle_T$ from both the variational VSCF approach of Section \ref{sec:vscf222} and the perturbative p-VPT approach of Section \ref{sec:pyropt}. 

The values of $\langle Q_{\textup{A}_1}\rangle_T$ for the p-VPT approach are obtained directly from Eq. (\ref{eq:avfin}) by restricting the sum over \textbf{${\bf q}$} points to just ${\bf q} = {\bf 0}$. For VSCF, the values of $\langle Q_{\textup{A}_1}\rangle_T$ are obtained according to Eq. (\ref{eq:uuu}) from a stochastic Metropolis-Hastings (MH) sampling of the vibrational Fock space described in Appendix \ref{app:MH}: full circles in Figure \ref{fig:Gamma}a) mark average values obtained for each temperature from 20 runs of the MH algorithm over $\tilde{N} =$ 5 million vibrational configurations each. The spread of computed values is reported in terms of their standard deviation as vertical error bars (not visible as in all cases they are within the size of the marker).

\begin{figure*}[t!]
    \centering
    \includegraphics[width=1.0\linewidth]{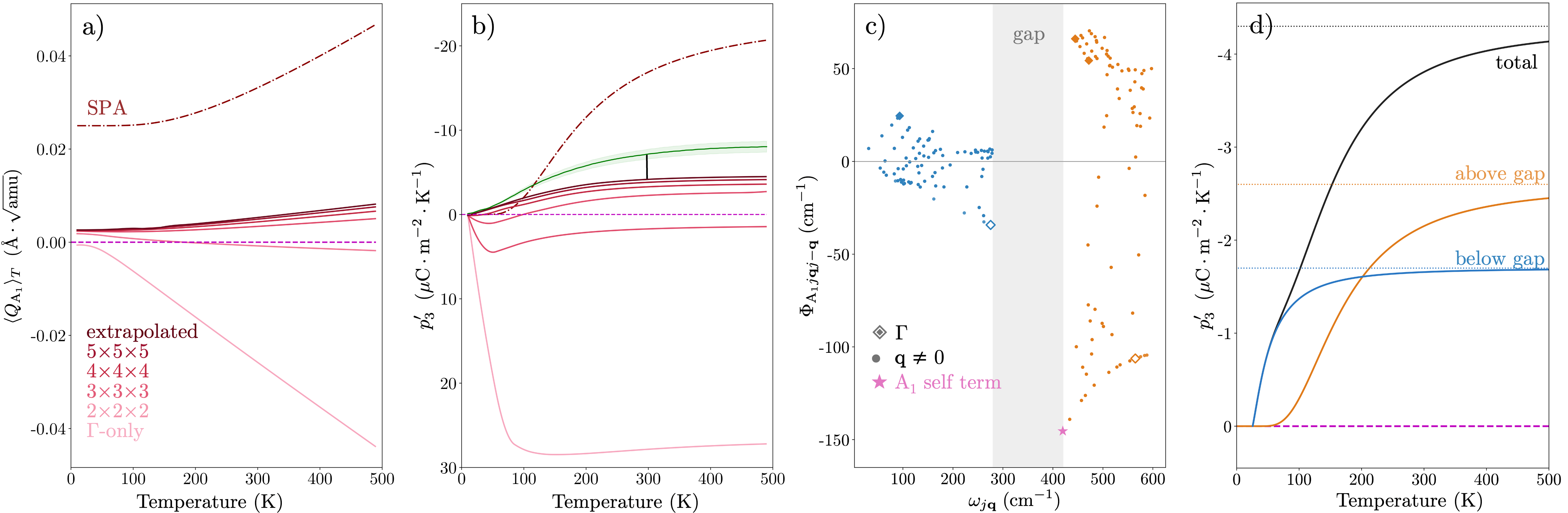}
    \caption{The effect of phonon dressing as a function of ${\bf q}$-mesh sampling in ZnO. a) Finite-temperature mean normal coordinate $\langle Q_{\textup{A}_1}\rangle_T$ as a function of temperature. b) Primary pyroelectric coefficient $p^\prime_3$ as a function of temperature. In a-b), results from the single phonon approximation are reported as a dash-dotted line. Results obtained by a progressive account for phonon dressing via a denser sampling of ${\bf q}$ points are reported as solid lines with progressively darker shades. In panel b), the green line represents the target primary pyroelectric coefficient $p_3^\prime = p_3 - p_3^{\prime\prime}$ of Figure \ref{fig:pyrexp}. c) Cubic coupling constants $\Phi_{\mathrm{A}_1 j\mathbf{q} j-\mathbf{q}}$ between the pyroelectrically active A$_1$(TO) mode and all other phonons, as a function of their harmonic frequency $\omega_{j\mathbf{q}}$. Phonons with $\mathbf{q}\neq0$ are shown as dots, colored blue below and orange above the frequency gap ($\sim$280--420~cm$^{-1}$, shaded). Phonons at $\Gamma$ are shown as diamonds: filled for doubly degenerate modes and empty for non-degenerate ones. The star marks the A$_1$(TO) self term $\Phi_{\mathrm{A}_1\mathrm{A}_1\mathrm{A}_1}$. d) Temperature dependence of the fully dressed primary pyroelectric coefficient $p_3'$ (black), decomposed into the contributions of phonons below (blue) and above (orange) the gap, as obtained from Eq.~(\ref{eq:pyroprimfinal}). Dotted lines give the corresponding classical ($k_\mathrm{B}T\gg\hbar\omega_{j\mathbf{q}}$) limits, in which each phonon contributes a temperature-independent amount proportional to $\Phi_{\mathrm{A}_1,j\mathbf{q},j-\mathbf{q}}/\omega_{j\mathbf{q}}$. The below-gap and total curves are shown only above 25~K: at lower temperatures the result is limited by the $\mathbf{q}$-mesh, which contains no phonons below 31~cm$^{-1}$.
    All values from the LDA.}
    \label{fig:dress}
\end{figure*}

The effect of phonon-phonon coupling at $\Gamma$ is large and such as to produce a qualitative change in the temperature dependence of $\langle Q_{\textup{A}_1}\rangle_T$, which increases with temperature at the SPA level and now decreases with temperature upon inclusion of phonon couplings at $\Gamma$. The two approaches, VSCF and p-VPT, provide a very similar and consistent description of the effect of such phonon couplings, but they do so at a significantly different computational cost. Indeed, the two approaches require different sets of anharmonic force constants $\Phi$: while in principle VSCF requires all anharmonic force constants among the set of modes at $\Gamma$, p-VPT only requires cubic force constants of type $\Phi_{\textup{A}_1 j{\bf 0} j{\bf 0}}$ according to Eq. (\ref{eq:avfin}). In practice, for our VSCF calculations we computed all cubic and quartic force constants involving up to two distinct modes. The inclusion of quartic terms is necessary to avoid unbounded mean-field potentials, as discussed in Section \ref{sec:BPA}. Furthermore, while p-VPT only requires force constants that involve the A$_1$ mode, VSCF also requires couplings among other modes. The two sets of anharmonic force constants needed for VSCF and p-VPT are reported in Figure \ref{fig:Gamma}b): they amount to 198 independent components for VSCF (81 cubic and 117 quartic) and just 9 for p-VPT.

Having established that the perturbative description closely matches the variational one but at a much lower cost, we therefore proceed to analyze further the effect of couplings with phonons with ${\bf q} \neq {\bf 0}$ just at the p-VPT level. We analyze the effect of phonon dressing beyond $\Gamma$ in Figure \ref{fig:dress} on both the finite-temperature mean normal coordinate $\langle Q_{\textup{A}_1}\rangle_T$ and the primary pyroelectric coefficient $p^\prime_3$, in a 0--500 K temperature range. The two quantities are computed from Eqs. (\ref{eq:avfin}) and (\ref{eq:pyroprimfinal}), respectively, where ${\bf q}$ points are sampled according to isotropic primitive meshes of increasing size: $1\times 1\times 1$ (i.e.\textbf{,} $\Gamma$ only), $2\times 2\times 2$, $3\times 3\times 3$, $4\times 4\times 4$, and $5\times 5\times 5$. In the figure, lines with a progressively darker shade of red correspond to progressively denser meshes. Both quantities in Figure \ref{fig:dress}a) and b) vary significantly as ${\bf q}$-sampling increases and eventually converge at $5\times 5\times 5$. The temperature dependence of $\langle Q_{\textup{A}_1}\rangle_T$ changes even qualitatively when passing from a $\Gamma$-only sampling to a fully converged ${\bf q}$-sampling: it steeply decreases with temperature at $\Gamma$ while it increases with temperature when computed from fully dressed phonons. Such flipping of the sign of $d\langle Q_{\textup{A}_1}\rangle_T/dT$ is then reflected in the sign of the pyroelectric coefficient. It is worth noting that both properties exhibit the same trend and sign when evaluated from fully dressed phonons as they do when computed from the SPA.

The fully dressed primary pyroelectric coefficient only slightly underestimates the target values -- green line in Figure \ref{fig:dress}b) -- derived from consolidated experimental values in Figure \ref{fig:pyrexp}. Let us note that this slight underestimate is fully consistent with a missing additional contribution to the pyroelectric coefficient of ZnO, estimated to be $\sim -4$ $\mu \text{C}\cdot \text{m}^{-2}\cdot \text{K}^{-1}$ at 300 K and marked by a vertical black line in Figure \ref{fig:dress}b), due to higher-than-linear order terms in the expansion of the spontaneous polarization in Eq. (\ref{eq:polexp}), also referred to as the electron-phonon renormalization contribution.\cite{liu2018mechanisms} While data reported in Figure \ref{fig:dress} refer to the LDA, very similar trends and values are obtained from PBE and PBE0 calculations, as reported in the SM, which shows that the primary pyroelectric coefficient in ZnO is rather insensitive to the particular choice of xc functional.

To further rationalize the effect of phonon dressing, we decompose the fully dressed
primary pyroelectric coefficient of Eq.~(\ref{eq:pyroprimfinal}) into individual phonon
contributions. Figure \ref{fig:dress}c) shows the cubic coupling constants
$\Phi_{\mathrm{A}_1,j\mathbf{q},j-\mathbf{q}}$ between the pyroelectrically
active A$_1$(TO) mode and all other phonons of the $5\times5\times5$
$\mathbf{q}$-mesh, as a function of their harmonic frequency, at LDA level. The phonon
spectrum of ZnO splits into two groups separated by a gap between
$\sim$280 and $\sim$420~cm$^{-1}$, as shown in Figure \ref{fig:bands}. Modes below the gap, which include the
acoustic and low-lying optical branches, couple only weakly to the A$_1$
mode ($|\Phi|\lesssim 30$~cm$^{-1}$). Modes above the gap couple
much more strongly, with values spread over both signs from about $-150$
to $+70$~cm$^{-1}$; the A$_1$ self term lies at the negative end
of this range. The small couplings below the gap are nonetheless far from
negligible. Eq. (\ref{eq:pyroprimfinal}) can be rewritten exactly as
$p_3'=\sum_{j\mathbf{q}} p_{3,j\mathbf{q}}'^{\,\mathrm{cl}}\,
C_\mathrm{E}(\hbar\omega_{j\mathbf{q}}/k_\mathrm{B}T)/k_\mathrm{B}$,
where $C_\mathrm{E}$ is the Einstein heat capacity of mode $j\mathbf{q}$ and
$p_{3,j\mathbf{q}}'^{\,\mathrm{cl}}\propto
\Phi_{\mathrm{A}_1,j\mathbf{q},j-\mathbf{q}}/\omega_{j\mathbf{q}}^2$ is its
temperature-independent classical contribution. The $1/\omega_{j\mathbf{q}}^2$
factor therefore strongly amplifies low-frequency modes. Because large
couplings of opposite sign partly cancel above the gap, the net
contributions of the two groups end up comparable: in their classical
limits the modes below and above the gap give $-1.70$ and
$-2.60$~$\mu$C\,m$^{-2}$\,K$^{-1}$, for a total of
$-4.30$~$\mu$C\,m$^{-2}$\,K$^{-1}$.


Because each mode enters weighted by its own Einstein function, the two
groups switch on at very different temperatures, see Figure \ref{fig:dress}d). Below
$\sim$100~K the response comes almost entirely from modes below the gap.
Their contribution rises steeply and reaches 90\% of its classical limit
by $\sim$150~K. In contrast, the optical modes above the gap are thermally
inactive up to $\sim$80~K, since their population is exponentially
suppressed for $\hbar\omega\gtrsim420$~cm$^{-1}$. The two contributions
become equal at $\sim$210~K, and above that temperature the optical
branches account for essentially all the remaining temperature dependence.
At room temperature, modes below the gap give
$-1.66$~$\mu$C\,m$^{-2}$\,K$^{-1}$ (44\% of the total) and modes above it
$-2.15$~$\mu$C\,m$^{-2}$\,K$^{-1}$ (56\%). At 500~K the total,
$-4.14$~$\mu$C\,m$^{-2}$\,K$^{-1}$, is within 4\% of its classical limit.
This picture accounts directly for the shortcomings of the SPA at low
temperature. The SPA retains only the A$_1$ mode at $\sim$420~cm$^{-1}$, so
it is exponentially suppressed exactly where the actual response is carried
entirely by phonons below the gap. In the fully dressed result, by contrast,
the A$_1$ self term contributes only $-0.13$~$\mu$C\,m$^{-2}$\,K$^{-1}$ at
300~K. Finally, this discussion shows that the heat-capacity-like temperature
dependence of $p_3'$ seen in Figure \ref{fig:pyrexp} is intrinsic to the p-VPT theory
of Eq. (\ref{eq:pyroprimfinal}). 


\section{Conclusions and Perspectives}

In summary, we have presented a  periodic extension of vibrational perturbation theory (p-VPT) specifically tailored for the evaluation of the primary (clamped-lattice) pyroelectric effect in polar solids. This allows for a systematic assessment of the effects of anharmonic phonon-phonon couplings at a modest cost. Indeed, this perturbative approach yields compact working expressions for the finite-temperature atomic displacements and primary pyroelectric coefficients, effectively bypassing the combinatorial complexity inherent to traditional variational methods like vibrational self-consistent field (VSCF) and vibrational configuration interaction (VCI). To address the challenges of the latter schemes, we have also devised a stochastic Metropolis-Hastings approach for sampling the vibrational Fock space. The methodology has been implemented in a developmental version of the \textsc{Crystal} electronic structure code.

Using bulk wurtzite-structured ZnO as a rigorous benchmark system, we demonstrated that p-VPT achieves remarkable quantitative agreement with variational calculations. Crucially, it does so at a fraction of the computational expense, requiring significantly fewer independent anharmonic force constants (e.g., merely 9 cubic components compared to the 198 cubic and quartic components required by VSCF at the \(\Gamma \) point). In particular, because it does not require even-order potential terms to remain stable -- in contrast to variational approaches, where truncation at odd orders may lead to unbounded total or mean-field potentials and thus to severe numerical instabilities -- p-VPT provides a significant practical simplification for the first-principles treatment of pyroelectricity in crystalline systems. Moreover, consistent results across local (LDA), gradient-corrected (PBE), and hybrid (PBE0) functionals demonstrate that both the physical trends and the accuracy of p-VPT are largely insensitive to the specific choice of exchange-correlation approximation, at least in the case of ZnO.

Our anharmonic investigation into the structural and dynamical contributions to the primary pyroelectric coefficient in ZnO highlights two key physical insights:
\begin{itemize}
\item We introduce the Single Phonon Approximation (SPA): restricting the anharmonic treatment strictly to the pyroelectrically active \(A_1(\text{TO})\) mode at \(\Gamma \) (i.e. neglecting phonon-phonon couplings {\bf and} the intrinsic anharmonicity of all other modes) yields a qualitatively sound description of the primary pyroelectric coefficient across a wide temperature range. Given its minimal computational cost, we think the SPA has the potential to become a powerful tool for fast, high-throughput screening in pyroelectric materials design and discovery. At the same time, its validity remains to be assessed for strongly anharmonic systems.

\item We analyze systematically the role of phonon dressing: incorporating phonon-phonon couplings between the pyroelectrically active mode at $\Gamma$ and other modes at ${\bf q} \neq {\bf 0}$ via dense Brillouin zone meshes is essential for achieving quantitative accuracy. Remarkably, we observe a qualitative inversion in the temperature dependence of the mean normal mode coordinate \(\langle Q_{A_1} \rangle_T\) when progressing from a \(\Gamma \)-only sampling to a fully converged \(5\times5\times5\) ${\bf q}$-mesh. Fully dressed phonons yield a primary pyroelectric coefficient that closely mirrors values inferred from consolidated experimental data. The remaining subtle underestimation is entirely consistent with the omission of higher-than-linear terms in the expansion of the spontaneous polarization.
\end{itemize}

Looking ahead, the efficiency of the p-VPT framework enables its immediate deployment to broader and structurally more complex families of materials, such as molecular/polymeric crystals, and perovskite ferroelectrics. The integration of higher-order electro-optical terms into the polarization expansion, also known as the electron-phonon renormalization contribution, would further improve the quantitative predictive power of the method. Finally, the high-throughput screening capabilities of the SPA could accelerate the atomistic design of pyroelectric materials for next-generation energy harvesting and catalysis.


\acknowledgments

This research has received funding from the Project CH4.0 under the MUR program ``Dipartimenti di Eccellenza 2023-2027'' (CUP: D13C22003520001). K.E. thanks Tekniikan edistämissäätiö (the Finnish Foundation for Technology Promotion) for funding (grant no. 12935). A.J.K. thanks Research Council of Finland for funding (grant no. 363733).

\appendix

\section{Second-Order Perturbative Treatment of Primary Pyroelectricity}
\label{app:pt12}

Let us start our second-order perturbative treatment of primary pyroelectricity by analyzing the integrals in Eqs. (\ref{eq:PTav0})--(\ref{eq:PTav2}), which are the zeroth-, first-, and second-order expressions, respectively, for the expectation value over state $\vert {\bf n} \rangle$ of the normal coordinate $\hat{Q}_{i{\bf 0}}$ of the $i$-th phonon at the Brillouin zone center. 

The zeroth-order term in Eq. (\ref{eq:PTav0}) can be worked out as follows: 
\begin{eqnarray}
   \langle {\bf n}^{(0)}\vert \hat{Q}_{i{\bf 0}}\vert {\bf n}^{(0)} \rangle  &=& 
  \langle {n_{i{\bf 0}}}^{(0)} \vert  \hat{Q}_{i{\bf 0}}\vert {n_{i{\bf 0}}}^{(0)}\rangle \prod_{j{\bf q}}\!\!\left.\right.^\prime \langle {n_{j{\bf q}}}^{(0)} \vert {n_{j{\bf q}}}^{(0)}\rangle \nonumber \\
   &=& \langle {n_{i{\bf 0}}}^{(0)} \vert  \hat{Q}_{i{\bf 0}}\vert {n_{i{\bf 0}}}^{(0)}\rangle = 0 \; ,
\end{eqnarray}
where the primed product runs over all phonons with $j{\bf q} \neq i{\bf 0}$, and where in the last step we have exploited the fact that, for any single quantum harmonic oscillator, the integral $\langle {n} \vert  \hat{Q}\vert {m}\rangle$ does not vanish only for $m = n \pm 1$, with:
\begin{eqnarray}
   \label{eq:monmodint1a}
   \langle {n} \vert  \hat{Q}\vert {n+1}\rangle &=& \sqrt{\hbar\frac{n + 1}{2 \omega}} \\
   \label{eq:monmodint1b}
   \langle {n} \vert  \hat{Q}\vert {n-1}\rangle &=& \sqrt{\hbar \frac{n}{2 \omega}} \; .
\end{eqnarray}

For the first-order term in Eq. (\ref{eq:PTav1}), we proceed as follows:
\begin{widetext}
\begin{eqnarray}
    \langle {\bf n}^{(1)}\vert \hat{Q}_{i{\bf 0}}\vert {\bf n}^{(0)} \rangle 
    &=& \sum_{\bf m}{'} 
    \frac{\braket{{\bf m}^{(0)} | \hat{H}^{(1)}| {\bf n}^{(0)}}}
    {E_{\bf n}^{(0)} - E_{\bf m}^{(0)}} 
    \braket{{\bf m}^{(0)}| \hat{Q}_{i{\bf 0}} | {\bf n}^{(0)}} \nonumber \\
    &=& \sum_{\bf m}{'} 
    \frac{\braket{{\bf m}^{(0)} | \hat{H}^{(1)}| {\bf n}^{(0)}}}
    {E_{\bf n}^{(0)} - E_{\bf m}^{(0)}} \left(\langle {m_{i{\bf 0}}}^{(0)} \vert  \hat{Q}_{i{\bf 0}}\vert {n_{i{\bf 0}}}^{(0)}\rangle \prod_{j{\bf q}}\!\!\left.\right.^\prime \langle {m_{j{\bf q}}}^{(0)} \vert {n_{j{\bf q}}}^{(0)}\rangle\right) \nonumber\\
    &=& \left[ \frac{\braket{({\bf n} + {\bf 1}_{i{\bf 0}})^{(0)} | \hat{H}^{(1)}| {\bf n}^{(0)}}}
    {E_{\bf n}^{(0)} - E_{{\bf n} + {\bf 1}_{i{\bf 0}}}^{(0)}} \langle (n_{i{\bf 0}}+1)^{(0)} \vert  \hat{Q}_{i{\bf 0}}\vert {n_{i{\bf 0}}}^{(0)}\rangle + \frac{\braket{({\bf n} - {\bf 1}_{i{\bf 0}})^{(0)} | \hat{H}^{(1)}| {\bf n}^{(0)}}}
    {E_{\bf n}^{(0)} - E_{{\bf n} - {\bf 1}_{i{\bf 0}}}^{(0)}} \langle (n_{i{\bf 0}}-1)^{(0)} \vert  \hat{Q}_{i{\bf 0}}\vert {n_{i{\bf 0}}}^{(0)}\rangle\right] \nonumber\\   
    &=& \left[ \frac{\braket{({\bf n} + {\bf 1}_{i{\bf 0}})^{(0)} | \hat{H}^{(1)}| {\bf n}^{(0)}}}
    {-\hbar \omega_{i{\bf 0}}} \sqrt{\hbar\frac{n_{i{\bf 0}} + 1}{2 \omega_{i{\bf 0}}}} + \frac{\braket{({\bf n} - {\bf 1}_{i{\bf 0}})^{(0)} | \hat{H}^{(1)}| {\bf n}^{(0)}}}
    {\hbar \omega_{i{\bf 0}}} \sqrt{\hbar\frac{n_{i{\bf 0}}}{2 \omega_{i{\bf 0}}}}\right] \nonumber\\ 
    \label{eccola}
    &=&  
      \frac{1 }{\sqrt{2\hbar\omega_{i{\bf 0}}^3}}\left[ \braket{({\bf n} \!-\! {\bf 1}_{i{\bf 0}})^{(0)} | \hat{H}^{(1)}| {\bf n}^{(0)}} \sqrt{n_{i{\bf 0}}}
    - \braket{({\bf n} \!+\! {\bf 1}_{i{\bf 0}})^{(0)} | \hat{H}^{(1)}| {\bf n}^{(0)}}
    \sqrt{{n_{i{\bf 0}} + 1}}     
     \right] \; ,
\end{eqnarray}
\end{widetext}
where in the first step we have used Eq. (\ref{eq:n1}) for the first-order correction to state $\vert {\bf n} \rangle$. In the third step the term within round brackets in the previous expression is evaluated, which does not vanish only if configurations \textbf{${\bf m}$} and \textbf{${\bf n}$} differ just by one quantum of excitation on phonon $i{\bf 0}$ while excitation quanta for all other phonons are the same. Thus, the sum over all vibrational configurations \textbf{${\bf m}$} reduces to just two configurations ${\bf n} + {\bf 1}_{i{\bf 0}}$ and ${\bf n} - {\bf 1}_{i{\bf 0}}$, where ${\bf 1}_{i{\bf 0}} = (0, \dots, 0, 1_{i{\bf 0}}, 0, \dots, 0)$ is a unit configuration vector with one quantum of excitation in mode $i{\bf 0}$ and zero in all other modes. In the fourth step, we exploit Eqs. (\ref{eq:monmodint1a}) and (\ref{eq:monmodint1b}).

With $H^{(1)}=\frac{1}{3!}\sum_{i{\bf q}\jq'\,k\bq''}\Phi_{i{\bf q}\jq'\,k\bq''}\hat Q_{i{\bf q}}\hat Q_{\jq'}\hat Q_{k\bq''}$, as per Eq.~(\ref{PesTaylor1}), a term connects $\bn$ to $\bn\pm\one$ only if it changes mode $\iz$ by one quantum and leaves every other mode unchanged.
Crystal-momentum conservation ($\bq+\bq'+\bq''=\mathbf{G}$) leaves two cases:
\begin{itemize}
\item[(i)] Self term (ST): all three indices equal $\iz$, corresponding to operator $\hat Q^3_{\iz}$, which appears once in the triple sum and thus has a weight of $\tfrac{1}{3!}\Phi_{\iz\iz\iz}$.
\item[(ii)] Mixed terms (MT): one index is $\iz$, the other two are $(\jq,\jmq)$ with $\jq\neq\iz$. As index $\iz$ can sit in any of the three slots, the weight of these terms is of $\tfrac{3}{3!}\Phic=\tfrac12\Phic$.
\end{itemize}
For the self term of case (i), we get the following expression for the quantity within squared brackets in Eq. (\ref{eccola}):
\begin{widetext}
\begin{equation}
\label{ppll}
\bra{n_{\iz}-1}\hat Q^3_{\iz}\ket{n_{\iz}}\sqrt{n_{\iz}}-\bra{n_{\iz}+1}\hat Q^3_{\iz}\ket{n_{\iz}}\sqrt{n_{\iz}+1}
=3\left(\frac{\hbar}{2\omega_{\iz}}\right)^{3/2}\big[n^2_{\iz}-(n_{\iz}+1)^2\big]
=-3\left(\frac{\hbar}{2\omega_{\iz}}\right)^{3/2}(2n_{\iz}+1) \; ,
\end{equation}
\end{widetext}
where we have exploited the following property:
\begin{eqnarray}
\bra{n-1}\hat Q^3\ket{n} &=& 3n\sqrt{n}\left(\frac{\hbar}{2\omega}\right)^{3/2} \\
\bra{n+1}\hat Q^3\ket{n} &=& 3(n+1)\sqrt{n+1}\left(\frac{\hbar}{2\omega}\right)^{3/2} \; .
\end{eqnarray}
Insertion of Eqs. (\ref{ppll}) into Eq. (\ref{eccola}) yields the following self term contribution to $\langle {\bf n}^{(1)}\vert \hat{Q}_{i{\bf 0}}\vert {\bf n}^{(0)} \rangle$:
\begin{eqnarray}
\langle {\bf n}^{(1)}\vert \hat{Q}_{i{\bf 0}}\vert {\bf n}^{(0)} \rangle_\textup{ST} &=& \frac{1}{6}\Phi_{\iz\iz\iz}\cdot\left(-3\right)(2n_{\iz}+1)\frac{\hbar}{4\omega^3_{\iz}} \nonumber \\
\label{unadidue}
&=&-\frac{\hbar}{2}\,\Phi_{\iz\iz\iz}\,\frac{2n_{\iz}+1}{4\omega_{\iz}\,\omega^2_{\iz}} \; .
\end{eqnarray}

\vspace{10pt}
\noindent For the mixed terms of case (ii), the matrix elements within squared brackets in Eq. (\ref{eccola}) factorize into a mode-$\iz$ part and a mode-$\jq$ part:
\begin{widetext}
\begin{eqnarray}
\bra{(n_{\iz}-1)}\Qiz\ket{n_{\iz}}\sqrt{n_{\iz}}-\bra{(n_{\iz}+1)}\Qiz\ket{n_{\iz}}\sqrt{n_{\iz}+1}
\label{due}
&=&\sqrt{\frac{\hbar}{2\omega_{\iz}}}\big[n_{\iz}-(n_{\iz}+1)\big]=-\sqrt{\frac{\hbar}{2\omega_{\iz}}} \; ; \\
\label{ecconoi}
\bra{n_{\jq}}\hat Q_{\jq}\hat Q_{\jmq}\ket{n_{\jq}}&=&\frac{\hbar}{\omega_{\jq}}\Big(n_{\jq}+\frac12\Big)
=\frac{\hbar\,(2n_{\jq}+1)}{2\omega_{\jq}} \; ,
\end{eqnarray}
\end{widetext}
where in Eq. (\ref{ecconoi}) we have exploited:
\begin{eqnarray}
   \label{eq:monmodint2a}
   \langle {n} \vert  \hat{Q}\hat{Q}^\dagger\vert {n}\rangle &=& \frac{\hbar}{\omega}\left(n + \frac{1}{2}\right) \; .
\end{eqnarray}
Insertion of Eqs. (\ref{due}) and (\ref{ecconoi}) into Eq. (\ref{eccola}) yields the following mixed terms contribution to $\langle {\bf n}^{(1)}\vert \hat{Q}_{i{\bf 0}}\vert {\bf n}^{(0)} \rangle$:
\begin{widetext}
\begin{equation}
\label{duedidue}
\langle {\bf n}^{(1)}\vert \hat{Q}_{i{\bf 0}}\vert {\bf n}^{(0)} \rangle_\textup{MT} = \sum_{j{\bf q} \neq {i{\bf 0}}}\frac{1}{\sqrt{2\hbar\omega^3_{\iz}}}\cdot\frac12\Phic\cdot\left(-\sqrt{\frac{\hbar}{2\omega_{\iz}}}\right)
\cdot\frac{\hbar(2n_{\jq}+1)}{2\omega_{\jq}}
= -\frac{\hbar}{2} \sum_{j{\bf q} \neq {i{\bf 0}}} \,\Phic\,\frac{2n_{\jq}+1}{4\omega_{\jq}\omega^2_{\iz}} \; ,
\end{equation}
\end{widetext}
where the sum runs over all phonons with $j{\bf q} \neq {i{\bf 0}}$ (apart from the three acoustic modes at $\Gamma$). Let us note that Eq. (\ref{duedidue})   has exactly the same form of Eq. (\ref{unadidue}) with $\jq=\iz$. Therefore the two cases merge into a single
sum over all phonons, including $\iz$ itself:
\begin{eqnarray}
\label{eq:ppp}
\bra{\bn^{(1)}}\Qiz\ket{\bn^{(0)}} &=&
\langle {\bf n}^{(1)}\vert \hat{Q}_{i{\bf 0}}\vert {\bf n}^{(0)} \rangle_\textup{ST} + \langle {\bf n}^{(1)}\vert \hat{Q}_{i{\bf 0}}\vert {\bf n}^{(0)} \rangle_\textup{MT} \nonumber \\
&=& -\frac{\hbar}{2}\sum_{\jq}\Phic\,\frac{2n_{\jq}+1}{4\omega_{\jq}\omega^2_{\iz}} . 
\end{eqnarray}

\vspace{10pt}
\noindent A similar derivation process can be applied to the evaluation of the two second-order terms in Eq. (\ref{eq:PTav2}), which we find to vanish:
\begin{eqnarray}
    \langle {\bf n}^{(1)}\vert {\hat{Q}}_{i{\bf 0}}\vert {\bf n}^{(1)} \rangle &=& 0 \; ,\\
    \langle {\bf n}^{(2)}\vert {\hat{Q}}_{i{\bf 0}}\vert {\bf n}^{(0)} \rangle &=& 0 \; .
\end{eqnarray}
We report the explicit derivation in the SM.\cite{ESIPYRO} It follows that the expectation value of the normal coordinate $\hat{Q}_{i{\bf 0}}$ over state $\vert {\bf n} \rangle$ reduces to the first-order contribution of Eq. (\ref{eq:ppp}):
\begin{eqnarray}
    \langle {\bf n}\vert \hat{Q}_{i{\bf 0}}\vert {\bf n} \rangle &\simeq& 2\lambda \langle {\bf n}^{(1)}\vert \hat{Q}_{i{\bf 0}}\vert {\bf n}^{(0)} \rangle \nonumber \\
    &\simeq& -\lambda \hbar \sum_{j{\bf q}} 
    \Phi_{i{\bf 0} j{\bf q} j-{\bf q}}
    \frac{2 n_{j{\bf q}} + 1}{4\omega_{j {\bf q}} \omega_{i {\bf 0}}^2} \; .
\end{eqnarray}
The thermal average, $\langle Q_{i{\bf 0}}\rangle_T$, of the normal coordinate ${Q}_{i{\bf 0}}$ can then be computed up to second order by combining Eq. (\ref{eq:av2}) with Eq. (\ref{eq:ppp}):
\begin{eqnarray}
\langle Q_{i{\bf 0}}\rangle_T &\simeq& 2\lambda\frac{1}{Z^{(0)}} \sum_{\bf n}   \langle {\bf n}^{(1)}\vert \hat{Q}_{i{\bf 0}}\vert {\bf n}^{(0)} \rangle  e^{-\beta E_{\bf n}^{(0)}} \nonumber \\
\label{eq:ooo}
&\simeq& - {\lambda \hbar} \sum_{j{\bf q}} 
    \frac{\Phi_{i{\bf 0} j{\bf q} j-{\bf q}}}{4\omega_{j {\bf q}} \omega_{i {\bf 0}}^2}
    \underbrace{\left[\frac{1}{Z^{(0)}}
\sum_{\bf n}
    (2 n_{j{\bf q}}\! +\! 1) e^{-\beta E_{\bf n}^{(0)}}\right]}_{X_{j{\bf q}}} \; . \nonumber \\
\end{eqnarray}
Let us recall that $E_{\bf n}^{(0)} = \sum_{i{\bf q}} \varepsilon_{i{\bf q},n_{i{\bf q}}}^{(0)}$. Thus, the multi-phonon harmonic partition function $Z^{(0)}$ can be factorized into single-phonon harmonic partition functions, $Z^{(0)}_{i{\bf q}}$, as follows:
\begin{eqnarray}
Z^{(0)} &=& \sum_{{\bf n}} e^{-\beta E_{\bf n}^{(0)}}
= \sum_{{\bf n}} e^{-\beta \sum_{i{\bf q}} \varepsilon_{i{\bf q},n_{i{\bf q}}}^{(0)}} \nonumber \\
&=& \prod_{i{\bf q}} \sum_{n_{i{\bf q}} =0}^\infty e^{-\beta  \varepsilon_{i{\bf q},n_{i{\bf q}}}^{(0)}} = \prod_{i{\bf q}} Z^{(0)}_{i{\bf q}} \; .
\end{eqnarray}
Thus, the term within square brackets in Eq. (\ref{eq:ooo}) can be written as:
\begin{eqnarray}
X_{j{\bf q}} &=& \frac{\prod_{l{\bf q}^\prime}\sum_{n_{l{\bf q}^\prime} =0}^\infty(2 n_{j{\bf q}}\! +\! 1)e^{-\beta  \varepsilon_{l{\bf q}^\prime,n_{l{\bf q}^\prime}}^{(0)}}}{\prod_{l{\bf q}^\prime}Z^{(0)}_{l{\bf q}^\prime}} \nonumber \\
\label{eq:aaa}
&=&\frac{\sum_{n_{j{\bf q}} =0}^\infty(2 n_{j{\bf q}}\! +\! 1)e^{-\beta  \varepsilon_{j{\bf q},n_{j{\bf q}}}^{(0)}}}{Z^{(0)}_{j{\bf q}}} \nonumber \\
&=& \frac{1 + e^{-\beta \hbar\omega_{j {\bf q}}}}{1 - e^{-\beta \hbar\omega_{j {\bf q}}}} \; ,
\end{eqnarray}
where we have used $\sum_{n=0}^\infty x^n = \frac{1}{1-x}$ and $\sum_{n=0}^\infty nx^n = \frac{x}{(1-x)^2}$. Insertion of Eq. (\ref{eq:aaa}) into Eq. (\ref{eq:ooo}), together with the per-unit-cell normalization factor $1/N_q$ and setting $\lambda = 1$, provides Eq. (\ref{eq:avfin}).

\section{Anharmonic Coupling Force Constants from Finite Differences}
\label{app:phi}

The perturbative treatment of pyroelectricity illustrated in Section \ref{sec:pyropt} and Appendix \ref{app:pt12} reveals that the only anharmonic force constants entering the primary pyroelectric coefficient up to second order in the perturbation are cubic ones of type $\Phi_{i{\bf 0} j{\bf q} j-{\bf q}}$, which can be conveniently expressed as follows:
\begin{eqnarray}
\label{eq:dercub}
\Phi_{i{\bf 0} j{\bf q} j-{\bf q}} &=& \left.\frac{\partial^3 E}{\partial {Q}_{i{\bf 0}} \partial {Q}_{j{\bf q}} \partial {Q}_{j-{\bf q}}}\right\vert_\textup{eq.} \nonumber \\
&=& \frac{\partial}{\partial {Q}_{i{\bf 0}}} \!\!\!\left.\left[ \frac{\partial^2 E}{\partial {Q}_{j{\bf q}} \partial {Q}_{j-{\bf q}}}\right]  \right\vert_\textup{eq.} \!\!\!\!= \left.\frac{\partial \lambda_{j{\bf q}}}{\partial {Q}_{i{\bf 0}}} \right\vert_\textup{eq.} \, ,
\end{eqnarray}
where $\lambda_{j{\bf q}}$ is the $j$-th eigenvalue of the ${\bf W}^{\bf q}$ dynamical matrix from Eq. (\ref{eq:diag}). We evaluate the derivative in the last expression of Eq. (\ref{eq:dercub}) numerically from finite differences by computing harmonic phonons at the equilibrium nuclear configuration ${\bf 0}$ and at an atomically displaced nuclear configuration obtained by moving the atoms along the $Q_{i{\bf 0}}$ normal coordinate by a step $\Delta Q_{i{\bf 0}}$ as follows:
\begin{equation}
   \Phi_{i{\bf 0} j{\bf q} j-{\bf q}} = \frac{\lambda_{j{\bf q}}({\bf 0} +\Delta Q_{i{\bf 0}}) - \lambda_{j{\bf q}}({\bf 0})}{\Delta Q_{i{\bf 0}}} \; ,
\end{equation}
where we have used a step of $\Delta Q_{i{\bf 0}} = $ 0.03 \AA{} $\cdot \,\sqrt{\text{amu}}$.   


\section{Finite-Temperature Properties from Metropolis-Hastings Sampling of the Vibrational Fock Space}
\label{app:MH}

We illustrate a Metropolis-Hastings (MH) stochastic approach\cite{metropolis1953equation,hastings1970monte} to the calculation of finite-temperature properties over vibrational states that can each be labeled by a vibrational configuration, $\vert s \rangle \equiv \vert {\bf n} \rangle$. In this case, the finite-temperature mean of an operator $\hat{\bullet}$ would be given by:  
\begin{equation}
    \label{eq:thermaverage2}
     \langle \hat{\bullet}\rangle_T = \sum_{\bf n}  \langle {\bf n}\vert\hat{\bullet}\vert {\bf n}\rangle p({\bf n};T) \quad \textup{with} \quad p({\bf n};T) = \frac{1}{Z} e^{-\beta E_{\bf n}} \; ,
\end{equation}
where $E_{\bf n}$ is the energy of state $\vert {\bf n}\rangle$, $p({\bf n};T)$ is its Boltzmann probability, and where $Z = \sum_{\bf m} e^{-\beta E_{\bf m}}$ is the probability distribution normalization factor in the usual form of the canonical partition function. In many practical circumstances -- including our VSCF description of anharmonic vibrational states discussed in Section \ref{sec:vscftheo} -- the normalization factor $Z$ is unknown as it cannot be converged by direct sum over a computationally affordable number of states, which makes Eq. (\ref{eq:thermaverage2}) of no straightforward use for the evaluation of thermal averages. We propose instead the use of a MH stochastic approach to select a set of $\tilde{N}$ configurations $\tilde{\bf n}$ so that the thermal average in Eq. (\ref{eq:thermaverage2}) can be computed from:
\begin{equation}
\label{eq:uuu2}
   \langle \hat{\bullet}\rangle_T \simeq \frac{1}{\tilde{N}} \sum_{{\tilde{\bf n}}} \langle {\tilde{\bf n}}\vert \hat{\bullet}\vert {\tilde{\bf n}} \rangle  \; .
\end{equation}
The MH algorithm indeed allows one to draw a sequence of samples (Markov chain) from an unknown probability distribution $p(x)$ if a function $f(x)$ proportional to the probability distribution is known and can be computed. The samples are accumulated iteratively starting from an initial sample $\tilde{x}^0$. Specifically, at each iteration $i \geq 1$, the algorithm proposes a candidate $x^c$ for the next sample based on the last sample from the previous iteration, $\tilde{x}^{i-1}$. Then, with some probability, the candidate is either accepted, $\tilde{x}^i = x^c$, or it is rejected, in which case the candidate value is discarded and $\tilde{x}^i = \tilde{x}^{i-1}$. At each iteration $i$, the probability of acceptance of a new candidate is determined by comparing the values of $f(x^c)$ and $f(\tilde{x}^{i-1})$, where, by construction, $\frac{f(x^c)}{f(\tilde{x}^{i-1})} \equiv \frac{p(x^c)}{p(\tilde{x}^{i-1})}$. The method used at each iteration to propose a new candidate from a previous sample depends on a {\it proposal probability function} $g(x\vert y)$ that represents the probability of proposing sample $x$ from sample $y$.

\begin{figure}[t!]
    \centering
    \includegraphics[width=1.0\linewidth]{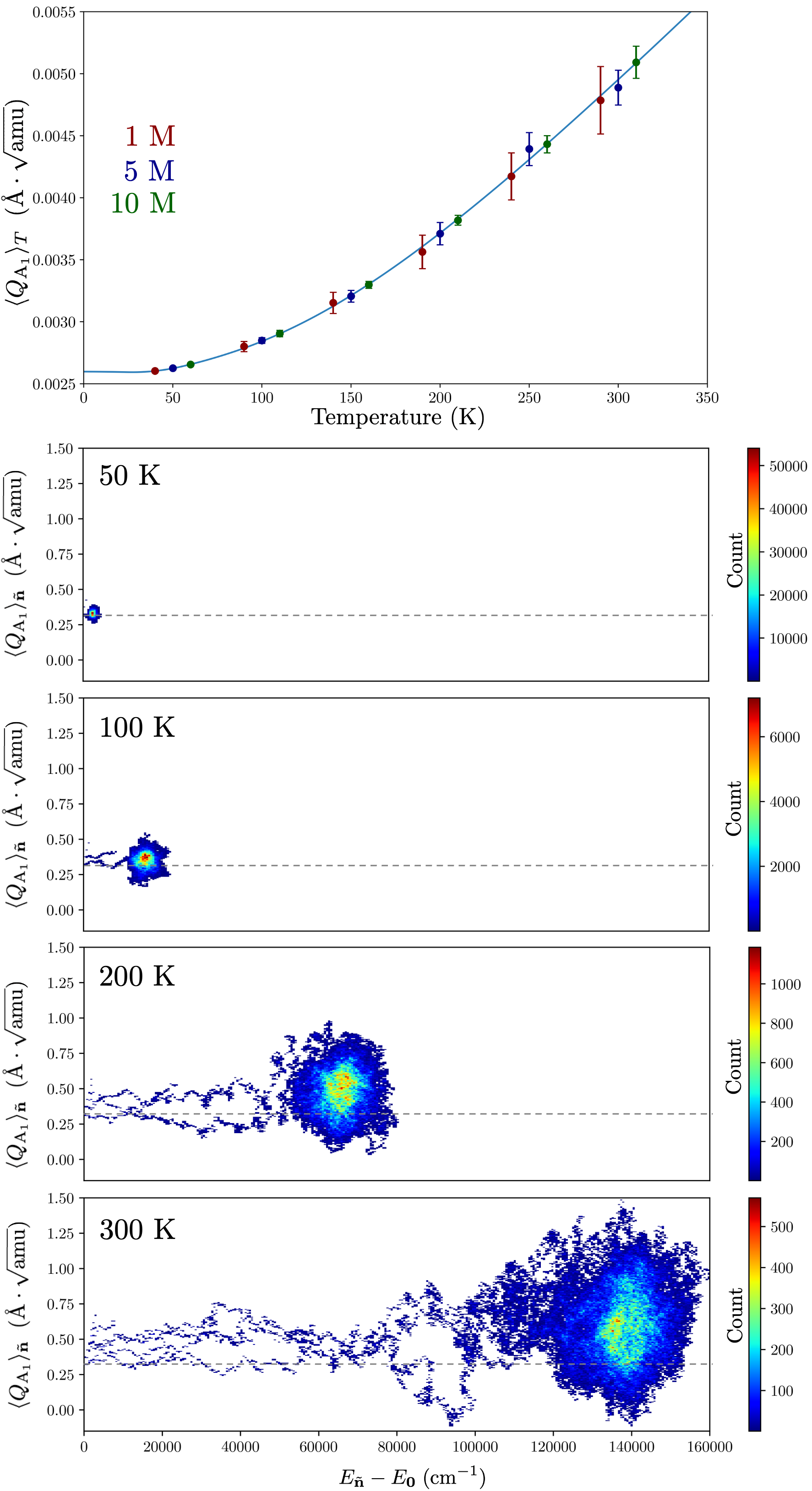}
    \caption{Stochastic MH approach for the calculation of the finite-temperature mean normal coordinate of the A$_1$ mode at $\Gamma$ of ZnO, {$\langle Q_{\textup{A}_1}\rangle_T$}. Top panel: {$\langle Q_{\textup{A}_1}\rangle_T$} as a function of temperature, as computed within the p-VPT approach. The solid line is obtained from the reference closed-form expression of Eq. (\ref{eq:avfin}), while symbols represent mean values obtained from the stochastic MH approach of Eq. (\ref{eq:uuu2}) for different values of $\tilde{N}$ (i.e., of sampled vibrational configurations), with $\tilde{N} = 1 \cdot 10^{6}, 5 \cdot 10^{6}, 1 \cdot 10^{7}$. The mean over twenty runs is reported in all cases. The spread of the obtained values over the twenty runs is reported in terms of standard deviation as a vertical bar in the figure. Lower panels: 2D histogram representation of the vibrational configurations sampled within the stochastic MH approach along a representative Markov chain at four different temperatures. All results refer to phonons computed within the LDA on a $5\times 5\times 5$ \textbf{${\bf q}$}-mesh.}
    \label{fig:MH}
\end{figure}

For our purpose, the unknown target probability distribution is $p({\bf n};T)$ from Eq. (\ref{eq:thermaverage2}) while the known function proportional to it is $f({\bf n};T) = e^{-\beta E_{\bf n}}$. We start the process from the ground state so that the initial sample is the null configuration vector $\tilde{{\bf n}}^0  = (0,0,0,\cdots,0,0,0)$. At each iteration, a new candidate configuration is proposed that is obtained from the previous one by either increasing or decreasing the quantum of excitation of a randomly picked phonon \textbf{$j{\bf q}$}: \textbf{$n^c_{j{\bf q}} = \tilde{n}^{i-1}_{j{\bf q}} \pm 1$}. Both the phonon whose excitation level is to be modified and the sign of its variation are determined randomly, unless the selected phonon has \textbf{$\tilde{n}^{i-1}_{j{\bf q}} = 0$}, in which case it can only be increased. Indeed, as phonon excitation quanta \textbf{$n_{j{\bf q}}$} cannot take negative values, the proposal probability function $g({\bf n}\vert {\bf m})$ is not symmetrical, i.e.\textbf{,} $g({\bf n}\vert {\bf m}) \neq g({\bf m}\vert {\bf n})$ for some pairs of configurations. For instance, if ${\bf n} = (0,0,0,\cdots,0,0,0)$ and ${\bf m} = (1,0,0,\cdots,0,0,0)$, the probability of proposing ${\bf m}$ from ${\bf n}$ is twice that of proposing ${\bf n}$ from ${\bf m}$: $\frac{g({\bf n}\vert {\bf m})}{g({\bf m}\vert {\bf n})} = \frac{1}{2}$. This leads to the following MH algorithm, where, for a given temperature $T$ and starting from the ground state configuration $\tilde{{\bf n}}^0$, at each iteration $i$ (starting from $i=1$):
\begin{enumerate}
    \item A random candidate configuration ${\bf n}^c$ is generated from the previous configuration $\tilde{{\bf n}}^{i-1}$ according to the proposal method described above;
    \item An acceptance probability $a({\bf n}^c, \tilde{{\bf n}}^{i-1})$ is computed as:
    \begin{equation}
      a({\bf n}^c, \tilde{{\bf n}}^{i-1}) = \textup{min}\left[ 1, \frac{f({\bf n}^c;T)}{f(\tilde{{\bf n}}^{i-1};T)} \cdot \frac{g(\tilde{{\bf n}}^{i-1}\vert {\bf n}^{c})}{g({\bf n}^c\vert \tilde{{\bf n}}^{i-1})}\right] \textbf{\; ,}
    \end{equation}
    \item A uniform random number $r \in [0,1]$ is generated and used to determine whether the candidate is accepted or rejected, according to:
    \begin{itemize}
    \item if $r \leq a({\bf n}^c, \tilde{{\bf n}}^{i-1})$ the candidate configuration is accepted and added to the sequence: $\tilde{{\bf n}}^{i}={{\bf n}}^c$;
    \item if $r > a({\bf n}^c, \tilde{{\bf n}}^{i-1})$ the candidate configuration is rejected and the previous configuration is copied forward: $\tilde{{\bf n}}^{i}=\tilde{{\bf n}}^{i-1}$. The total number of times a given configuration appears in the chain (its initial acceptance plus any consecutive rejections) determines its \textit{multiplicity} $m_{\tilde{{\bf n}}}$;
    \end{itemize}
    \item Set $i = i+1$ and go back to step 1 unless $i \geq \tilde{N}$ in which case exit.
\end{enumerate}

The computational cost of Eq. (\ref{eq:uuu2}) can be reduced by summing only over {\it unique} configurations $\tilde{{\bf n}}_u$ within the Markov chain and accounting for their multiplicity:
\begin{equation}
\label{eq:uuu3}
   \langle \hat{\bullet}\rangle_T \simeq \frac{1}{\tilde{N}} \sum_{{\tilde{\bf n}}_u} m_{\tilde{{\bf n}}_u}\langle {\tilde{\bf n}}_u\vert \hat{\bullet}\vert {\tilde{\bf n}}_u \rangle  \; , 
\end{equation}
where $\tilde{N} = \sum_{{\tilde{\bf n}}_u} m_{\tilde{{\bf n}}_u}$.

We test our MH stochastic sampling approach on the calculation of the finite-temperature mean normal coordinate of the A$_1$ mode at $\Gamma$ of ZnO (i.e.\textbf{,} $\hat{\bullet} = {\hat{Q}}_{\textup{A}_1}$) from our perturbative approach. This is because, within the p-VPT approach, {$\langle Q_{\textup{A}_1}\rangle_T$ can be evaluated both from Eq. (\ref{eq:uuu2}) with the MH approach, and from Eq. (\ref{eq:thermaverage2}), where it takes the closed-form reference expression reported in Eq. (\ref{eq:avfin}). We assess the MH approach with the help of Figure \ref{fig:MH}. In the top panel, we report the finite-temperature mean normal coordinate $\langle Q_{\textup{A}_1}\rangle_T$ as a function of temperature, as derived from anharmonic phonons computed within the LDA on a $5\times 5\times 5$ \textbf{${\bf q}$}-mesh. Reference values obtained from the closed-form expression of Eq. (\ref{eq:avfin}) are reported as a solid line. The same quantity has been computed from Eq. (\ref{eq:uuu2}) with the MH approach for different values of $\tilde{N}$ (i.e.\textbf{,} of sampled vibrational configurations), with $\tilde{N} = 1 \cdot 10^{6}$, $5 \cdot 10^{6}$, $1 \cdot 10^{7}$ at six different temperatures. For each batch size and temperature, the stochastic MH approach is repeated twenty times. The means, $\langle\langle Q_{\textup{A}_1}\rangle_T\rangle$, over the twenty runs are reported in the figure as symbols. The spread of the obtained values over the twenty runs is reported in terms of standard deviation as a vertical bar in the figure. Several trends emerge: i) at low temperature, the results obtained from the MH approach are perfectly superimposed {on} the reference ones for all batch sizes and show negligible spread; ii) as expected, the spread increases with temperature and decreases with batch size; iii) the higher the temperature, the larger the number of sampled configurations needed for an accurate description. A setup with {a} batch size of 10 million configurations and twenty repetitions (green symbols in the figure) is found to provide an excellent description over the explored temperature range, 0--350 K.    

We analyze further the MH stochastic sampling of the vibrational Fock space with the help of the lower panels of Figure \ref{fig:MH}, where we visually inspect a single representative Markov chain for $\tilde{N} = 1 \cdot 10^{6}$ at four different temperatures, 50, 100, 200, and 300 K. A 2D histogram representation is used to highlight the density of sampled vibrational configurations as a function of two properties of each sampled configuration ${\tilde{\bf n}}$: its energy $E_{\tilde{\bf n}}$ and its mean normal mode coordinate $\langle Q_{\textup{A}_1}\rangle_{\tilde{\bf n}} \equiv \langle {\tilde{\bf n}} \vert \hat{Q}_{\textup{A}_1} \vert {\tilde{\bf n}} \rangle$. The dashed horizontal line in the panels marks the mean normal mode coordinate of the vibrational ground state configuration ${\tilde{\bf n}} = {\bf 0}$ and serves as a guide to the eye. As temperature increases, three distinctive effects are clearly highlighted in the figure: configurations with higher energy are progressively being sampled, with a progressively larger spread, and, most importantly for the description of pyroelectricity, with a progressively larger mean normal coordinate $\langle Q_{\textup{A}_1}\rangle_{\tilde{\bf n}}$. Figure \ref{fig:MH} provides a quantitative assessment of the three effects.

\bibliography{DatabaseBIBLIO}
\bibliographystyle{apsrev}

\end{document}


\maketitle

\section{Phonon Band Structure of ZnO (Harmonic)}

\subsection{Effect of xc Approximation}

\begin{figure}[h!!]
    \centering
    \includegraphics[width=0.9\linewidth]{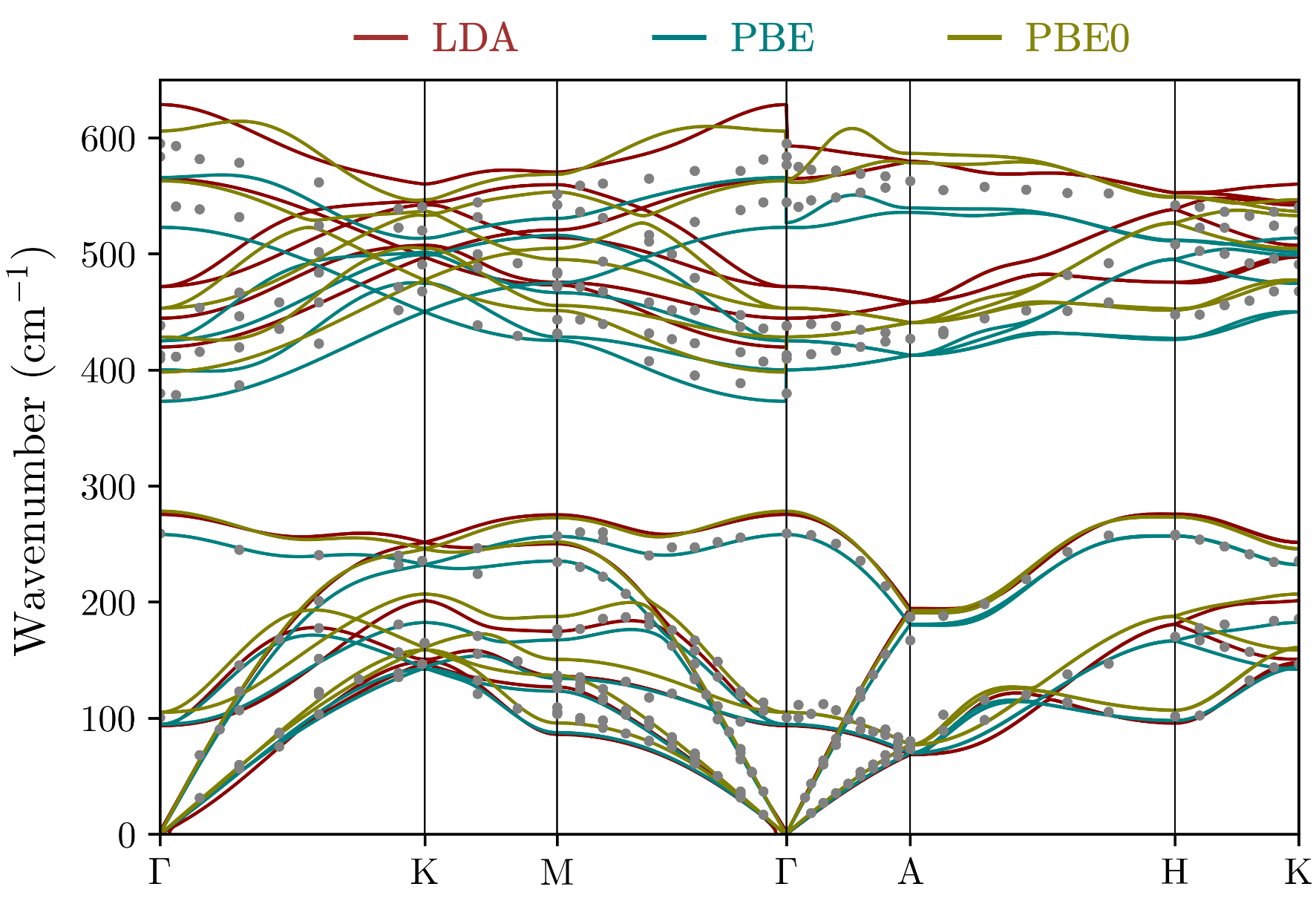}
    \caption{Harmonic phonon band structure of bulk ZnO along the $\Gamma$--K--M--$\Gamma$--A--H--K path, obtained by Fourier interpolation of force constants computed with the frozen-phonon approach on a $5\times5\times5$ supercell, including the non-analytical correction for long-range electrostatic interactions. Dark red, teal and olive solid lines correspond to LDA, PBE and PBE0, respectively. Gray dots are experimental data.\cite{serrano2010phonon, hewat1970lattice, thoma1974lattice, serrano2003dependence}}
    \label{fig:phonon_band_ESI}
\end{figure}

Fig. \ref{fig:phonon_band_ESI} compares the harmonic phonon dispersion of bulk ZnO computed with the LDA, PBE and PBE0 xc functionals against experimental data.\cite{serrano2010phonon, hewat1970lattice, thoma1974lattice, serrano2003dependence}
Force constants were obtained with the frozen-phonon approach on a $5\times5\times5$ supercell and Fourier interpolated along the $\Gamma$--K--M--$\Gamma$--A--H--K path, with the non-analytical correction for long-range electrostatic interactions capturing the LO--TO splitting at $\Gamma$ included. All three functionals reproduce the overall topology of the dispersion, including the gap between the low-frequency (below about 280~cm$^{-1}$) and high-frequency (above about 370~cm$^{-1}$) modes. The acoustic branches and the low-lying optical branches below about 200~cm$^{-1}$ are described accurately by all functionals and differ from one another by only a few cm$^{-1}$. Differences are larger for the optical modes where frequencies are slightly underestimated by PBE and slightly overestimated by LDA and PBE0.  

\subsection{Convergence with Supercell Size}

\begin{figure}[h!!]
    \centering
    \includegraphics[width=0.9\linewidth]{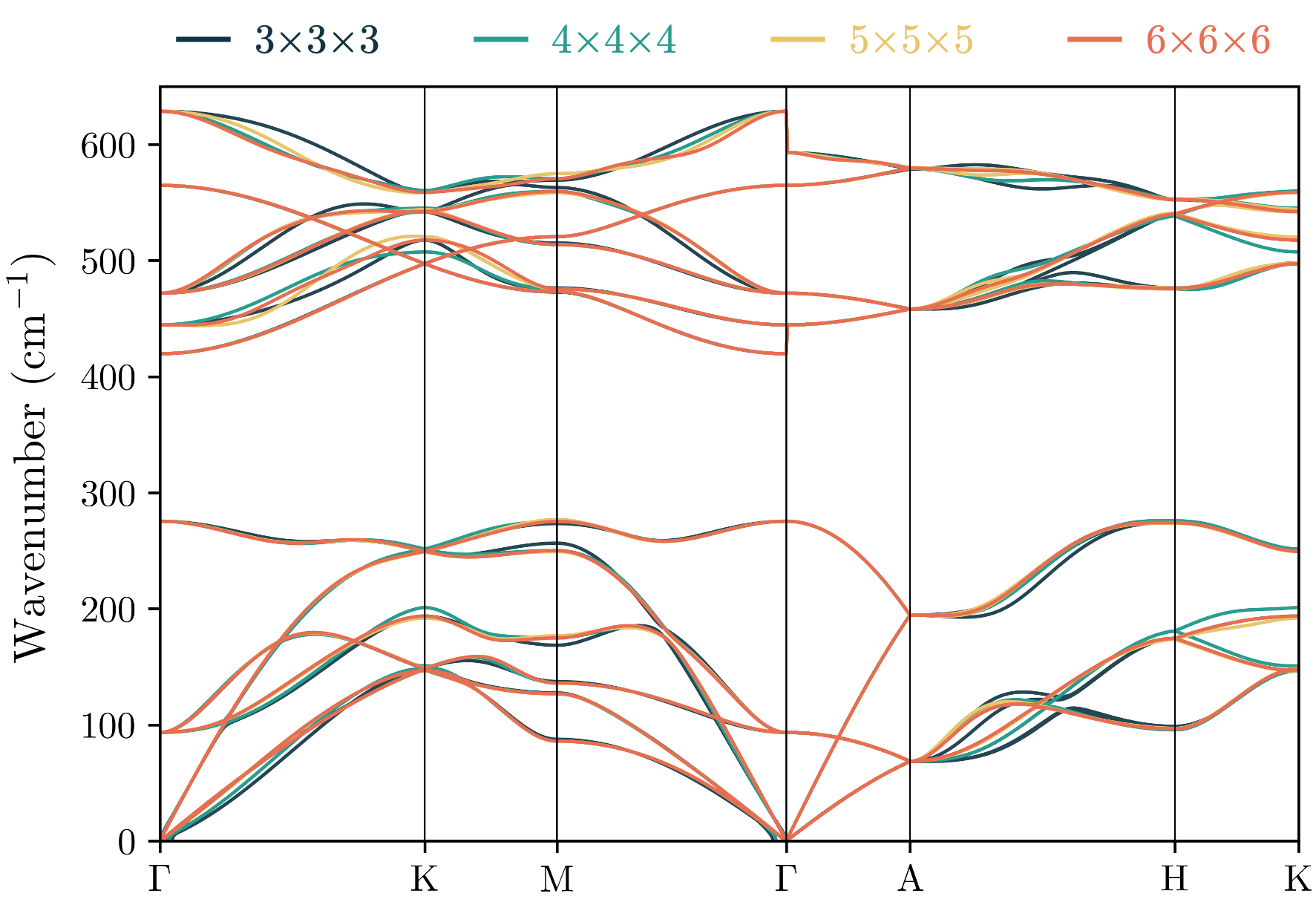}
    \caption{Convergence of the harmonic phonon band structure of bulk ZnO along the $\Gamma$--K--M--$\Gamma$--A--H--K path with respect to the supercell size  used in the frozen-phonon calculation of the force constants, computed at the LDA level. Dark blue, teal, yellow and orange solid lines correspond to $3\times3\times3$ (108 atoms), $4\times4\times4$ (256 atoms), $5\times5\times5$ (500 atoms) and $6\times6\times6$ (864 atoms) supercells, respectively. The non-analytical correction for long-range electrostatic interactions is included in all cases.}
    \label{fig:phonon_band_sc_conv}
\end{figure}

Fig. \ref{fig:phonon_band_sc_conv} shows the convergence of the harmonic phonon dispersion of bulk ZnO with respect to the size of the supercell used to compute the force constants within the direct frozen-phonon approach, from $3\times3\times3$ to $6\times6\times6$, at the LDA level. Table \ref{tab:kmesh} reports the $\bf k$-point meshes used in the electronic calculations for each supercell.  
As expected, frequencies at $\Gamma$ and A high-symmetry points are identical from all calculations, as they are commensurate to all explored supercells. The largest differences appear at the zone boundary and along the A--H and H--K lines, mostly for the low-frequency optical branches between about 100 and 200~cm$^{-1}$ and for the high-frequency branches around 500~cm$^{-1}$, where the $3\times3\times3$ and $4\times4\times4$ supercells deviate from the larger ones by up to about 10--20~cm$^{-1}$. The $5\times5\times5$ and $6\times6\times6$ results are virtually indistinguishable throughout the Brillouin zone, which supports the use of the $5\times5\times5$ supercell for all the calculations reported in the main text of this work.

\begin{table}[h!]
\caption{$\bf k$-point mesh and number of irreducible $\bf k$-point adopted in the electronic calculations for each (super)cell.}
\label{tab:kmesh}
\begin{tabular}{ccc}
\hline\hline
&& \\
Cell size & $\bf k$-point mesh & irreducible $\bf k$-points (IBZ)\\
&& \\
Primitive  &  $12 \times 12 \times 12$ & 133 \\ 
$2 \times 2 \times 2$  & $6 \times 6 \times 6$ & 34 \\
$3 \times 3 \times 3$  & $4 \times 4 \times 4$ & 13 \\
$4 \times 4 \times 4$  & $3 \times 3 \times 3$ & 7 \\
$5 \times 5 \times 5$  & $3 \times 3 \times 3$ & 7 \\
$6 \times 6 \times 6$  & $2 \times 2 \times 2$ & 4 \\
&&\\
\hline\hline  
\end{tabular}
\end{table}

\section{Primary Pyroelectric Coefficient of ZnO}

\subsection{Single Phonon Approximation}

\begin{figure}[h!!]
    \centering
    \includegraphics[width=1.0\linewidth]{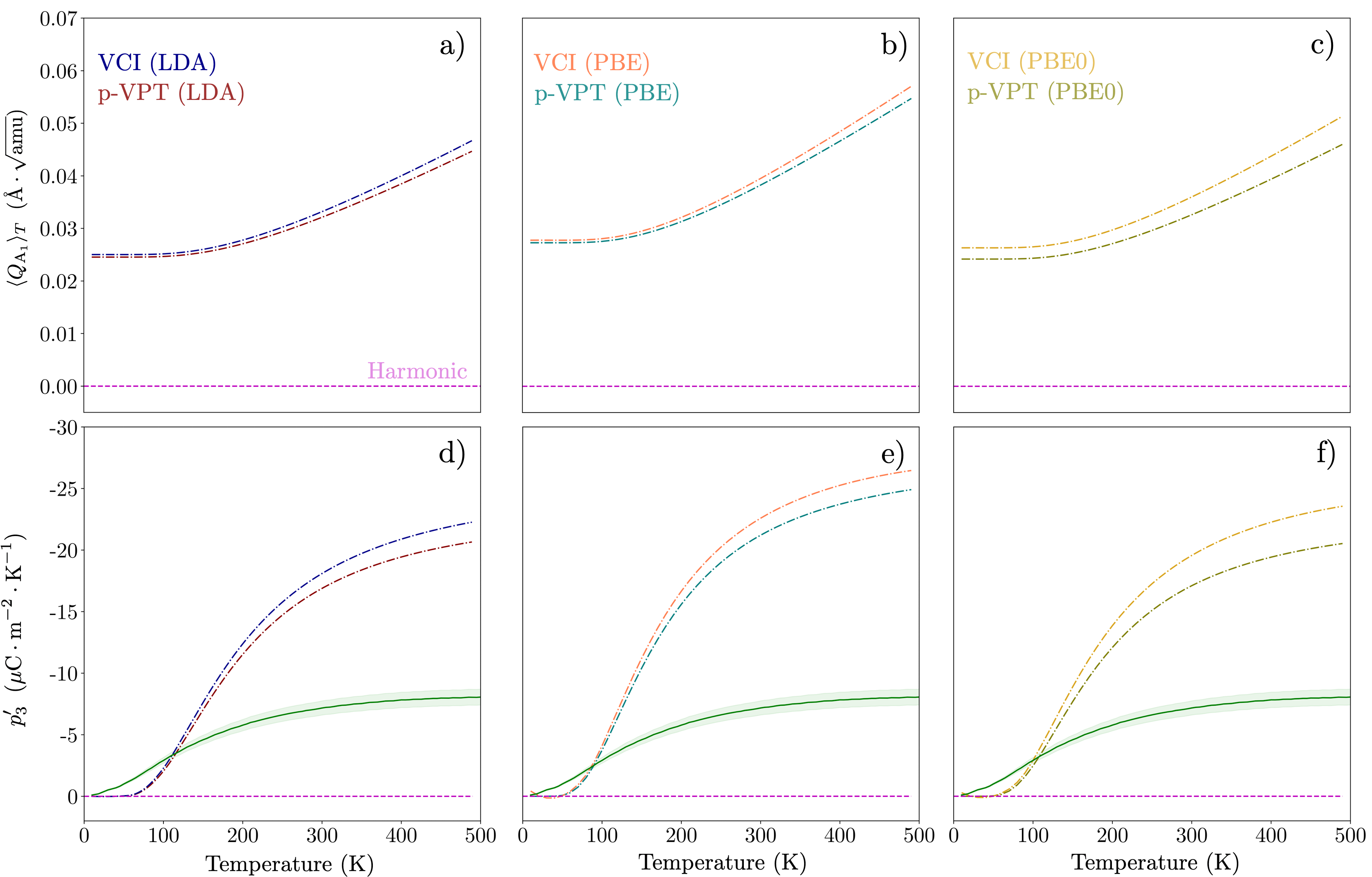}
    \caption{Temperature dependence of (a--c) the finite-temperature normal coordinate of the polar A$_1$ mode at $\Gamma$ of bulk ZnO, $\langle Q_{\mathrm{A}_1}\rangle_T$, and (d--f) the primary pyroelectric coefficient $p'_3$ of bulk ZnO obtained from the single phonon approximation, computed with (a,d) LDA, (b,e) PBE, (c,f) PBE0. In each panel, the two dash-dotted lines correspond to VCI and p-VPT results. The magenta horizontal dashed line corresponds to the harmonic description, for which both $\langle Q_{\mathrm{A}_1}\rangle_T$ and $p'_3$ are trivially null. The green solid line and associated shaded area correspond to target values deduced from consolidated experimental data and to a nominal uncertainty of $\pm 8\%$.\cite{heiland1966pyroelectricity, gong2026ab}} 
    \label{fig:BPA_xc_ESI}
\end{figure}

Fig. \ref{fig:BPA_xc_ESI} reports the finite-temperature  normal coordinate of the polar A$_1$ mode at $\Gamma$ of bulk ZnO, $\langle Q_{\text{A}_1}\rangle_T$ (panels a--c), and the primary pyroelectric coefficient of bulk ZnO, $p'_3$ (panels d--f) computed from the single phonon approximation (SPA), as obtained from the VCI and p-VPT approaches with the LDA, PBE and PBE0 xc functionals. 

All three functionals give a finite zero-point displacement of about 0.024--0.028~\AA$\sqrt{\text{amu}}$, which stays nearly constant up to about 100~K and then grows almost linearly with temperature. Accordingly, $p'_3$ is negligible below about 50~K, increases steeply in magnitude between 100 and 300~K, and approaches saturation at higher temperatures, reaching about $-22$ (LDA), $-26$ (PBE) and $-24$~$\mu$C~m$^{-2}$~K$^{-1}$ (PBE0) at 500~K at the VCI level. The perturbative p-VPT approach reproduces the VCI results closely: it underestimates $|p'_3|$ at 500~K by about 7\% for LDA and 6\% for PBE, and by about 13\% for PBE0, for which the two approaches also differ slightly in the zero-point value of $\langle Q_{\mathrm{A}_1}\rangle_T$. The dependence on the functional is therefore moderate, and the agreement between VCI and p-VPT holds for all three functionals.

\subsection{Fully Dressed Phonons}

\begin{figure}[h!!]
    \centering
    \includegraphics[width=0.8\linewidth]{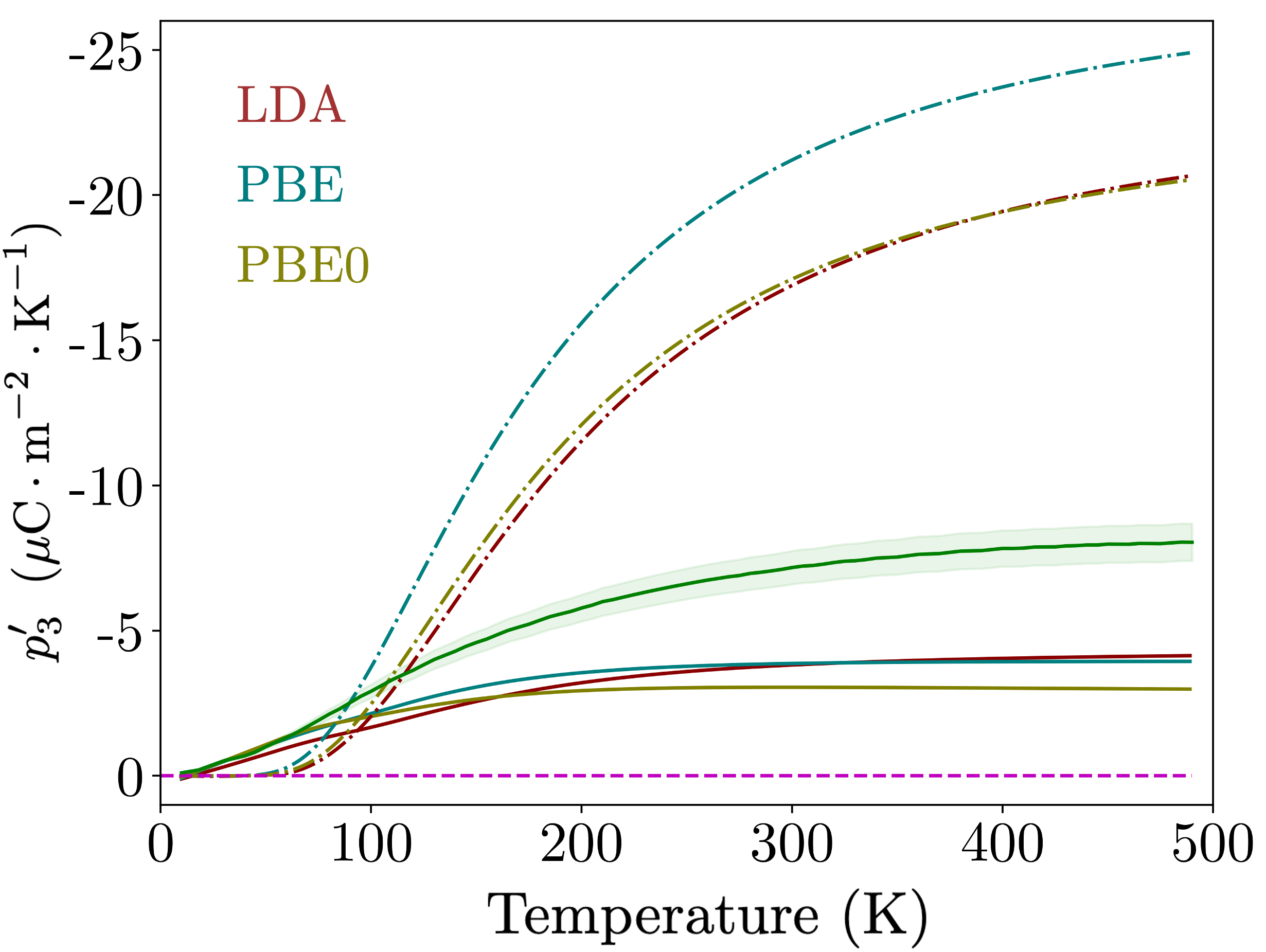}
    \caption{Temperature dependence of the primary pyroelectric coefficient $p'_3$ of bulk ZnO computed with the p-VPT approach from a $5\times5\times5$ {\bf q}-mesh from LDA (dark red), PBE (teal) and PBE0 (olive green). Dash-dotted and solid lines correspond to the SPA and fully dressed descriptions, respectively. The magenta dashed horizontal line corresponds to the harmonic description. The green solid line and associated shaded area correspond to target values deduced from consolidated experimental data and to a nominal uncertainty of $\pm8\%$, respectively.\cite{heiland1966pyroelectricity, gong2026ab}}
    \label{fig:pyro_xc_dressing_ESI}
\end{figure}

Fig. \ref{fig:pyro_xc_dressing_ESI} compares the primary pyroelectric coefficient $p'_3$ obtained from the p-VPT approach within the SPA with that obtained when phonon dressing is fully included, for the LDA, PBE and PBE0 xc functionals. Phonon dressing strongly reduces the magnitude of the coefficient: at 500~K, $p'_3$ goes from about $-21$ (LDA), $-25$ (PBE) and $-21$~$\mu$C~m$^{-2}$~K$^{-1}$ (PBE0) in the SPA to about $-4$ (LDA and PBE) and $-3$~$\mu$C~m$^{-2}$~K$^{-1}$ (PBE0) in the fully dressed description. 

Dressing also changes the low-temperature behavior qualitatively. In the SPA, $p'_3$ is negligible below about 60~K and then rises steeply. With full dressing, $p'_3$ increases in magnitude approximately linearly from the lowest temperatures and closely follows the target value up to about 100~K, before saturating above about 200~K. All three functionals show the same qualitative behavior. PBE and PBE0 agree up to about 150~K, after which PBE0 saturates earlier, while LDA rises more slowly but reaches the largest magnitude above about 350~K. The dependence on the xc functional is much smaller than the effect of phonon dressing.

\section{On the Vanishing Second-Order p-VPT Contributions to the Mean Normal Coordinate}

In the main text, the finite-temperature normal coordinate is evaluated up to first-order in the perturbative expansion of $\langle\mathbf{n}|\hat Q_{i0}|\mathbf{n}\rangle$, where the only non-vanishing contribution is $\langle\mathbf{n}^{(1)}|\hat Q_{i0}|\mathbf{n}^{(0)}\rangle$, see Eq. (38). In passing from first- to second-order, two additional contributions appear, $\langle\mathbf{n}^{(2)}|\hat Q_{i0}|\mathbf{n}^{(0)}\rangle$ and $\langle\mathbf{n}^{(1)}|\hat Q_{i0}|\mathbf{n}^{(1)}\rangle$. In the following, we show that both are null, so that the first-order expression of the main text is exact up to second order.

\subsection{The $\langle{\bf n}^{(2)}|\hat Q_{i{\bf 0}}|{\bf n}^{(0)}\rangle$ Term}

The second-order correction to the state, $|{\bf n}^{(2)}\rangle$, is given by Eq. (21) of the main text. Inserting it into $\langle{\bf n}^{(2)}|\hat Q_{i{\bf 0}}|{\bf n}^{(0)}\rangle$ yields:
\begin{align}
\label{eq:n2Qn0}
\langle{\bf n}^{(2)}|\hat Q_{i{\bf 0}}|{\bf n}^{(0)}\rangle
&= \sum_{\bf m}{}'
\frac{\langle{\bf m}^{(0)}|\hat H^{(2)}|{\bf n}^{(0)}\rangle}
     {E^{(0)}_{\bf n}-E^{(0)}_{\bf m}}
\langle{\bf m}^{(0)}|\hat Q_{i{\bf 0}}|{\bf n}^{(0)}\rangle \nonumber\\
&\quad+\sum_{\bf m}{}'\sum_{{\bf m}'}{}'
\frac{\langle{\bf m}^{(0)}|\hat H^{(1)}|{\bf m}'^{(0)}\rangle
      \langle{\bf m}'^{(0)}|\hat H^{(1)}|{\bf n}^{(0)}\rangle}
     {(E^{(0)}_{\bf n}-E^{(0)}_{\bf m})(E^{(0)}_{\bf n}-E^{(0)}_{{\bf m}'})}
\langle{\bf m}^{(0)}|\hat Q_{i{\bf 0}}|{\bf n}^{(0)}\rangle \nonumber\\
&= \sum_{\bf m}{}'
\frac{\langle{\bf m}^{(0)}|\hat H^{(2)}|{\bf n}^{(0)}\rangle}
     {E^{(0)}_{\bf n}-E^{(0)}_{\bf m}}
\left(\langle m_{i{\bf 0}}^{(0)}|\hat Q_{i{\bf 0}}|n_{i{\bf 0}}^{(0)}\rangle
\prod_{j{\bf q}}{}'\langle m_{j{\bf q}}^{(0)}|n_{j{\bf q}}^{(0)}\rangle\right) \nonumber\\
&\quad+\sum_{\bf m}{}'\sum_{{\bf m}'}{}'
\frac{\langle{\bf m}^{(0)}|\hat H^{(1)}|{\bf m}'^{(0)}\rangle
      \langle{\bf m}'^{(0)}|\hat H^{(1)}|{\bf n}^{(0)}\rangle}
     {(E^{(0)}_{\bf n}-E^{(0)}_{\bf m})(E^{(0)}_{\bf n}-E^{(0)}_{{\bf m}'})}
\left(\langle m_{i{\bf 0}}^{(0)}|\hat Q_{i{\bf 0}}|n_{i{\bf 0}}^{(0)}\rangle
\prod_{j{\bf q}}{}'\langle m_{j{\bf q}}^{(0)}|n_{j{\bf q}}^{(0)}\rangle\right) \nonumber\\
&= \frac{\langle({\bf n}+{\bf 1}_{i{\bf 0}})^{(0)}|\hat H^{(2)}|{\bf n}^{(0)}\rangle}{-\hbar\omega_{i{\bf 0}}}
\sqrt{\hbar\frac{n_{i{\bf 0}}+1}{2\omega_{i{\bf 0}}}}
+\frac{\langle({\bf n}-{\bf 1}_{i{\bf 0}})^{(0)}|\hat H^{(2)}|{\bf n}^{(0)}\rangle}{\hbar\omega_{i{\bf 0}}}
\sqrt{\hbar\frac{n_{i{\bf 0}}}{2\omega_{i{\bf 0}}}} 
\nonumber \\
&\quad+\sum_{\bf m}{}'\sum_{{\bf m}'}{}'
\frac{\langle{\bf m}^{(0)}|\hat H^{(1)}|{\bf m}'^{(0)}\rangle
      \langle{\bf m}'^{(0)}|\hat H^{(1)}|{\bf n}^{(0)}\rangle}
     {(E^{(0)}_{\bf n}-E^{(0)}_{\bf m})(E^{(0)}_{\bf n}-E^{(0)}_{{\bf m}'})}
\left(\langle m_{i{\bf 0}}^{(0)}|\hat Q_{i{\bf 0}}|n_{i{\bf 0}}^{(0)}\rangle
\prod_{j{\bf q}}{}'\langle m_{j{\bf q}}^{(0)}|n_{j{\bf q}}^{(0)}\rangle\right) 
\end{align}
The first two terms in the last expression of Eq. (\ref{eq:n2Qn0}) involve the second-order Hamiltonian, $\hat H^{(2)}$. Each quartic term in $\hat H^{(2)}$ is a product of four phonon position operators, each of which changes the corresponding phonon occupation number by $\pm1$; hence $\hat H^{(2)}|{\bf n}^{(0)}\rangle$ only contains harmonic states whose total number of quanta differs from that of $|{\bf n}^{(0)}\rangle$ by an even integer (0, $\pm2$ or $\pm4$). Since $|({\bf n}\pm{\bf 1}_{i{\bf 0}})^{(0)}\rangle$ differs from $|{\bf n}^{(0)}\rangle$ by a single quantum, orthogonality of the harmonic basis gives $\langle({\bf n}\pm{\bf 1}_{i{\bf 0}})^{(0)}|\hat H^{(2)}|{\bf n}^{(0)}\rangle=0$.

As per the last term in Eq. (\ref{eq:n2Qn0}), the factor $\langle m_{i{\bf 0}}^{(0)}|\hat Q_{i{\bf 0}}|n_{i{\bf 0}}^{(0)}\rangle \prod_{j{\bf q}}{}'\langle m_{j{\bf q}}^{(0)}|n_{j{\bf q}}^{(0)}\rangle$ restricts the sum over ${\bf m}$ to ${\bf m}={\bf n}\pm{\bf 1}_{i{\bf 0}}$, so that it reads:
\begin{align}
\label{eq:n2Qn0_H1}
\sum_{{\bf m}'}{}'
\frac{\langle{\bf m}'^{(0)}|\hat H^{(1)}|{\bf n}^{(0)}\rangle}
     {E^{(0)}_{\bf n}-E^{(0)}_{{\bf m}'}}
\left[
\frac{\langle({\bf n}+{\bf 1}_{i{\bf 0}})^{(0)}|\hat H^{(1)}|{\bf m}'^{(0)}\rangle}
     {E^{(0)}_{\bf n}-E^{(0)}_{{\bf n}+{\bf 1}_{i{\bf 0}}}}
\langle(n_{i{\bf 0}}+1)^{(0)}|\hat Q_{i{\bf 0}}|n_{i{\bf 0}}^{(0)}\rangle
\right. \nonumber\\
\qquad\qquad\left.
+\frac{\langle({\bf n}-{\bf 1}_{i{\bf 0}})^{(0)}|\hat H^{(1)}|{\bf m}'^{(0)}\rangle}
     {E^{(0)}_{\bf n}-E^{(0)}_{{\bf n}-{\bf 1}_{i{\bf 0}}}}
\langle(n_{i{\bf 0}}-1)^{(0)}|\hat Q_{i{\bf 0}}|n_{i{\bf 0}}^{(0)}\rangle
\right] \nonumber\\
\end{align}
Since each cubic term in $\hat H^{(1)}$ is a product of three phonon position operators, $\hat H^{(1)}$ only connects harmonic states whose total number of quanta differs by an odd integer. A non-zero product $\langle{\bf m}'^{(0)}|\hat H^{(1)}|{\bf n}^{(0)}\rangle \langle({\bf n}\pm{\bf 1}_{i{\bf 0}})^{(0)}|\hat H^{(1)}|{\bf m}'^{(0)}\rangle $ in Eq. (\ref{eq:n2Qn0_H1}) would thus require $|{\bf m}'^{(0)}\rangle$ to differ by an odd number of quanta from both $|{\bf n}^{(0)}\rangle$ and $|({\bf n}\pm{\bf 1}_{i{\bf 0}})^{(0)}\rangle$, which is impossible since the latter differ by a single quantum. 

Hence:
\begin{equation}
    \langle{\bf n}^{(2)}|\hat Q_{i{\bf 0}}|{\bf n}^{(0)}\rangle = 0 .
\end{equation}

\subsection{The $\langle{\bf n}^{(1)}|\hat Q_{i{\bf 0}}|{\bf n}^{(1)}\rangle$ Term}

We now turn to $\langle{\bf n}^{(1)}|\hat Q_{i{\bf 0}}|{\bf n}^{(1)}\rangle$. Since only $\hat H^{(1)}$ enters the first-order correction $|{\bf n}^{(1)}\rangle$, and the factor $\langle m_{i{\bf 0}}^{(0)}|\hat Q_{i{\bf 0}}|m_{i{\bf 0}}'^{(0)}\rangle \prod_{j{\bf q}}{}'\langle m_{j{\bf q}}^{(0)}|m_{j{\bf q}}'^{(0)}\rangle$ restricts the sum over ${\bf m}$ to ${\bf m}={\bf m}'\pm{\bf 1}_{i{\bf 0}}$, we obtain:
\begin{align}
\label{eq:n1Qn1}
\langle{\bf n}^{(1)}|\hat Q_{i{\bf 0}}|{\bf n}^{(1)}\rangle
&= \sum_{\bf m}{}'\sum_{{\bf m}'}{}'
\frac{\langle{\bf n}^{(0)}|\hat H^{(1)}|{\bf m}^{(0)}\rangle
      \langle{\bf m}'^{(0)}|\hat H^{(1)}|{\bf n}^{(0)}\rangle}
     {(E^{(0)}_{\bf n}-E^{(0)}_{\bf m})(E^{(0)}_{\bf n}-E^{(0)}_{{\bf m}'})}
\langle{\bf m}^{(0)}|\hat Q_{i{\bf 0}}|{\bf m}'^{(0)}\rangle \nonumber\\
&= \sum_{\bf m}{}'\sum_{{\bf m}'}{}'
\frac{\langle{\bf n}^{(0)}|\hat H^{(1)}|{\bf m}^{(0)}\rangle
      \langle{\bf m}'^{(0)}|\hat H^{(1)}|{\bf n}^{(0)}\rangle}
     {(E^{(0)}_{\bf n}-E^{(0)}_{\bf m})(E^{(0)}_{\bf n}-E^{(0)}_{{\bf m}'})}
\left(\langle m_{i{\bf 0}}^{(0)}|\hat Q_{i{\bf 0}}|m_{i{\bf 0}}'^{(0)}\rangle
\prod_{j{\bf q}}{}'\langle m_{j{\bf q}}^{(0)}|m_{j{\bf q}}'^{(0)}\rangle\right) \nonumber\\
&= \sum_{{\bf m}'}{}'
\frac{\langle{\bf m}'^{(0)}|\hat H^{(1)}|{\bf n}^{(0)}\rangle}
     {E^{(0)}_{\bf n}-E^{(0)}_{{\bf m}'}}
\left[
\frac{\langle{\bf n}^{(0)}|\hat H^{(1)}|({\bf m}'+{\bf 1}_{i{\bf 0}})^{(0)}\rangle}
     {E^{(0)}_{\bf n}-E^{(0)}_{{\bf m}'+{\bf 1}_{i{\bf 0}}}}
\langle(m'_{i{\bf 0}}+1)^{(0)}|\hat Q_{i{\bf 0}}|m_{i{\bf 0}}'^{(0)}\rangle
\right. \nonumber\\
&\qquad\qquad\left.
+\frac{\langle{\bf n}^{(0)}|\hat H^{(1)}|({\bf m}'-{\bf 1}_{i{\bf 0}})^{(0)}\rangle}
     {E^{(0)}_{\bf n}-E^{(0)}_{{\bf m}'-{\bf 1}_{i{\bf 0}}}}
\langle(m'_{i{\bf 0}}-1)^{(0)}|\hat Q_{i{\bf 0}}|m_{i{\bf 0}}'^{(0)}\rangle
\right]
\end{align}
As above, $\hat H^{(1)}$ only connects harmonic states whose total number of quanta differs by an odd integer. A non-zero product $\langle{\bf n}^{(0)}|\hat H^{(1)}|({\bf m}'\pm{\bf 1}_{i{\bf 0}})^{(0)}\rangle \langle{\bf m}'^{(0)}|\hat H^{(1)}|{\bf n}^{(0)}\rangle$ in Eq. (\ref{eq:n1Qn1}) would thus require both $|{\bf m}'^{(0)}\rangle$ and $|({\bf m}'\pm{\bf 1}_{i{\bf 0}})^{(0)}\rangle$ to differ by an odd number of quanta from $|{\bf n}^{(0)}\rangle$, which is impossible since the latter differ by a single quantum. 

Hence:
\begin{equation}
\langle{\bf n}^{(1)}|\hat Q_{i{\bf 0}}|{\bf n}^{(1)}\rangle = 0 .
\end{equation}

\section{CRYSTAL Input Deck}

We report below an example \textsc{Crystal} input deck for the calculation of the phonon band structure of bulk ZnO from the frozen-phonon approach, with a $5\times 5\times 5$ supercell.

{
  \setlength{\baselineskip}{12pt}
\begin{verbatim}
PBE0 - scelphono 5x5x5
CRYSTAL
0 0 0
186
3.26569491     5.21136084
2
30     3.333333333333E-01 -3.333333333333E-01  5.919269170278E-05
8      3.333333333333E-01 -3.333333333333E-01  3.824408073083E-01
SCELPHONO
5 0 0
0 5 0
0 0 5
FREQCALC
NUMDERIV
2
DISPERSI
BANDS
12 100
6
0 0 0  4 4 0
4 4 0  6 0 0
6 0 0  0 0 0
0 0 0  0 0 6
0 0 6  4 4 6
4 4 6  4 4 0
END
END
30 12
0 0 8 2.0 1.0
  405924.31028      .22442017483E-03
  60846.955735      .17402086626E-02
  13847.343092      .90513339565E-02
  3919.6158551      .36817341445E-01
  1276.3594167      .12004850256
  458.67254435      .28576057621
  178.28725246      .41087462062
  70.612192837      .21816962456
0 0 4 2.0 1.0
  443.88077950     -.24934274984E-01
  137.55875267     -.11817955766
  22.268083479      .55367318468
  9.5217310606      .52628934936
0 0 2 2.0 1.0
  14.874114065     -.22929955254
  2.4647517612      .71135484742
0 0 1 2.0 1.0
  1.0113272238       1.0000000000
0 0 1 0.0 1.0
  0.40               1.0000000000
0 1 1 0.0 1.0
  0.16               1.0 1.0
0 2 6 6.0 1.0
  2205.3508534      .23356240448E-02
  522.35300699      .19031022634E-01
  167.73055542      .89955758675E-01
  62.670045373      .26113248631
  25.109749456      .42348448173
  10.225142681      .24618926885
0 2 3 6.0 1.0
  40.713442521     -.30029667592E-01
  5.6247090696      .55575254864
  2.2279949116      .95581013442
0 2 1 0.0 1.0
  0.83354741691      1.0000000000
0 3 4 10.0 1.0
  88.554315311      .12728170015E-01
  25.721525557      .79394499843E-01
  9.1278367624      .24491506805
  3.4312364064      .40390526479
0 3 1 0.0 1.0
  1.2308920645      .40158491145
0 3 1 0.0 1.0
  0.39031845112      .21579805034
8 7
0 0 6 2.0 1.0
  27032.382631      0.21726302465E-03
  4052.3871392      0.16838662199E-02
  922.32722710      0.87395616265E-02
  261.24070989      0.35239968808E-01
  85.354641351      0.11153519115
  31.035035245      0.25588953961
0 0 2 2.0 1.0
  12.271113873      0.39768730901
  4.9159842006      0.24627849430
0 0 1 0.0 1.0
 0.90086482370      1.0000000000
0 1 1 0.0 1.0
 0.25000000000      1.0 1.0
0 2 4 4.0 1.0
  75.300554155      0.60685103418E-02
  17.743733858      0.41912575824E-01
  5.5355828651      0.16153841088
  2.0685535103      0.35706951311
0 2 1 0.0 1.0
 0.78238772422      1.0000000000
0 3 1 0.0 1.0
 1.2000000000       1.0000000000
99 0
ENDBAS
DFT
PBE0
ENDDFT
SHRINK
3 3
TOLINTEG
8 8 8 8 16
END
\end{verbatim}
}

\bibliography{MT/DatabaseBIBLIO}